\documentclass[letterpaper,11pt,fleqn]{article}
\pdfoutput=1
\usepackage[margin=1.2in]{geometry} %
\usepackage[utf8]{inputenc}
\usepackage{amsmath}
\usepackage{amssymb} 
\usepackage{mathtools}
\usepackage{booktabs} 
\usepackage{colortbl} 
\usepackage{dsfont} 
\usepackage[small]{caption} 
\usepackage{subcaption}
\usepackage{booktabs}
\usepackage{mathrsfs}	
\usepackage{cite} 
\usepackage{cancel}
\usepackage{nicefrac}
\usepackage{pifont} 
\usepackage[normalem]{ulem}
\usepackage[pdftex,hidelinks,colorlinks=true,citecolor=magenta,linkcolor=blue,urlcolor=blue]{hyperref}
\hypersetup{
    pdftitle = {Modular flavored dark matter},
    pdfauthor = {Baur, Chen, Knapp-Perez, Ramos-Sanchez}
}
\usepackage{cleveref} 
\usepackage{braket}
\usepackage{xfrac}
\usepackage{wrapfig}
\usepackage{microtype}
\usepackage[nolist]{acronym}
\usepackage{float}
\usepackage[compat=1.1.0]{tikz-feynman}
\usepackage{feynmp-auto}
\usepackage{amsfonts,stackengine,graphicx}
\usepackage{mwe}
\usepackage{bbm}
\usepackage{xspace}

\usepackage[all,cmtip]{xy}
\newlength{\xywd}
\newcommand{\xyrightarrow}[2][]{%
  \sbox{0}{$\scriptstyle#1$}%
  \xywd=\wd0
  \sbox{0}{$\scriptstyle#2$}%
  \ifdim\wd0>\xywd \xywd=\wd0 \fi
  \xymatrix@C\dimexpr\xywd+1em\relax{{}\ar[r]^{#2}_{#1}&{}}%
}

\usepackage[textsize=tiny,backgroundcolor=yellow,textwidth=80pt]{todonotes}

\setstackgap{S}{-1.5pt}

\makeatletter
\g@addto@macro\bfseries{\boldmath}
\makeatother
\DeclareMathOperator{\re}{Re}
\DeclareMathOperator{\im}{Im}

\newcommand{\D}{\mathrm{d}}
\newcommand{\I}{\mathrm{i}}
\newcommand{\e}{\mathrm{e}}
\newcommand{\Id}{\mathbbm{1}}

\newcommand{\SU}[1]{\ensuremath{\mathrm{SU}(#1)}}
\newcommand{\SL}[1]{\ensuremath{\mathrm{SL}(#1)}}
\newcommand{\U}[1]{\ensuremath{\mathrm{U}(#1)}}
\newcommand{\Z}[1]{\ensuremath{\mathbbm{Z}_{#1}}} 

\newcommand{\CP}{\ensuremath{\mathcal{CP}}\xspace}
\newcommand{\x}{\ensuremath{\times}}

\newcommand{\rep}[2][]{\ensuremath{{\boldsymbol{#2}_{#1}}}}

\allowdisplaybreaks

\UseCollection{xfrac}{plainmath}

\begin{document}

\newcommand\mytitle{Modular flavored dark matter} 

\begin{titlepage}
\begin{flushright}
UCI-TR-2024-12
\\
TUM-HEP 1520/24
\end{flushright}

\vspace*{2em}

\begin{center}
{\LARGE\sffamily\bfseries\mytitle}

\vspace{2em}

\renewcommand*{\thefootnote}{\fnsymbol{footnote}}

\textbf{%
Alexander Baur$^{a,b}$\footnote{alexander.baur@tum.de}, 
Mu--Chun Chen$^{c,}$\footnote{muchunc@uci.edu}, 
V.~Knapp--P\'erez$^{c,}$\footnote{vknapppe@uci.edu},\\
Sa\'ul Ramos--S\'anchez$^{a,}$\footnote{ramos@fisica.unam.mx},}
\\[8mm]
\textit{$^a$\small
~Instituto de F\'isica, Universidad Nacional Aut\'onoma de M\'exico, Cd.~de M\'exico C.P.~04510, M\'exico}
\\[5mm]
\textit{$^b$\small
~Physik Department, Technische Universit\"at M\"unchen,
James-Franck-Straße 1, 85748 Garching, Germany}
\\[5mm]
\textit{$^c$\small
~Department of Physics and Astronomy, University of California, Irvine, CA 92697-4575 USA
}
\end{center}

\vspace*{5mm}

\begin{abstract}
Discrete flavor symmetries have been an appealing approach for explaining the observed flavor structure, which is not justified in the \ac{SM}. Typically, these models require a so-called flavon field in order to give rise to the flavor structure upon the breaking of the flavor symmetry by the \ac{VEV} of the flavon. Generally, in order to obtain the desired vacuum alignment, a flavon potential that includes additional so-called driving fields is required. On the other hand, allowing the flavor symmetry to be modular leads to a structure where the couplings are all holomorphic functions that depend only on a complex modulus, thus greatly reducing the number of parameters in the model. We show that these elements can be combined to simultaneously explain the flavor structure and \ac{DM} relic abundance. We present a modular model with flavon vacuum alignment that allows for realistic flavor predictions while providing a successful fermionic \ac{DM} candidate.
\end{abstract}

\end{titlepage}
\renewcommand*{\thefootnote}{\arabic{footnote}}
\setcounter{footnote}{0}


\section{Introduction}
\label{sec:Introduction}

The origin of the mass hierarchies and mixings among the three generations of fermions is unexplained in the \ac{SM}. One possible  solution to address this so-called flavor puzzle
is the introduction of {\it traditional flavor symmetries} that allow for transformations among fermions of different flavors. 
These symmetries are independent of moduli and are known to provide explanations for the flavor structure of both the lepton and quark
sectors~\cite{Altarelli:2010gt,Ishimori:2010au,Hernandez:2012ra,King:2013eh,King:2014nza,King:2017guk,Feruglio:2019ybq}.
To achieve these results, models endowed with traditional flavor symmetries require the introduction
of two kinds of extra \ac{SM} singlet scalars: i) flavons, whose \ac{VEV}s are responsible for the non-trivial flavor structure of fermions,
and ii) driving fields~\cite{Altarelli:2005yx,Feruglio:2009iu,Ding:2013hpa,Li:2013jya,Muramatsu:2016bda,King:2019tbt,CarcamoHernandez:2019eme,Chen:2021prl}
that help to shape a suitable potential for obtaining the desired \ac{VEV} pattern. The traditional flavor
symmetry framework based on non-Abelian discrete symmetries has also proven helpful for \ac{DM}, where the flavor
symmetry plays the role of a stabilizer symmetry~\cite{Hirsch:2010ru}. Furthermore, (not necessarily discrete)
flavor symmetries can be successfully combined to explain both \ac{DM} and flavor anomalies
from \ac{SM} decays~\cite{He:2024iju,Acaroglu:2023phy,Acaroglu:2021qae}.

Another promising approach to explain the flavor parameters without introducing many scalars is provided by so-called modular flavor symmetries~\cite{Feruglio:2017spp,Criado:2018thu,Kobayashi:2018vbk,deAnda:2018ecu,Okada:2018yrn,Ding:2019xna,Kobayashi:2019xvz,Asaka:2019vev,Liu:2019khw,Ding:2023htn}. This approach has produced
good fits for leptons~\cite{Kobayashi:2018scp,Kobayashi:2018vbk,Novichkov:2018nkm,Novichkov:2018ovf,Novichkov:2018yse,Penedo:2018nmg,Criado:2019tzk,Ding:2019xna,Ding:2019zxk,Kobayashi:2019mna,Liu:2019khw,Ding:2020msi,Novichkov:2020eep,Wang:2020lxk,Li:2021buv}
and quarks~\cite{Okada:2018yrn,Kobayashi:2019rzp,Ding:2020zxw,Lu:2019vgm,King:2020qaj,Liu:2020akv,Okada:2020ukr,Chen:2021zty,Ding:2021eva,Zhao:2021jxg,Arriaga-Osante:2023wnu,Kikuchi:2023jap,Kikuchi:2023dow}.
In this scenario, instead of depending on flavon \ac{VEV}s, Yukawa couplings are replaced
by multiplets of modular forms depending on the half-period ratio $\tau$, which can be
considered a complex modulus. The basic problem then reduces to explaining why
$\tau$ stabilizes at its best-fit value $\langle\tau\rangle$, 
for which being close to the symmetry-enhanced points $\tau = \I, \; \e^{\nicefrac{2\pi \I}{3}}, \;  \I\infty$ can be advantageous~\cite{Feruglio:2021dte,Feruglio:2022koo,Petcov:2022fjf,Abe:2023ilq,Feruglio:2023mii}. Although this scheme has the potential to avoid the need for any
additional scalar field, the presence of flavons in addition to the modulus can be useful too (see, e.g.,\ Model 1 of~\cite{Feruglio:2017spp} and~\cite{Criado:2018thu}). Similarly to the traditional flavor case, modular flavor symmetries can also
serve as stabilizer symmetries for \ac{DM} candidates~\cite{Nomura:2019jxj,Behera:2020lpd,Hutauruk:2020xtk,Kobayashi:2021ajl,Zhang:2021olk}

Modular flavor symmetries arise naturally in top-down constructions, such as magnetized extra dimensions~\cite{Kobayashi:2016ovu,Kobayashi:2018rad,Kobayashi:2018bff,Kariyazono:2019ehj,Hoshiya:2020hki,Kikuchi:2020frp,Kikuchi:2020nxn,Ohki:2020bpo,Almumin:2021fbk,Kikuchi:2021ogn,Tatsuta:2021deu,Kikuchi:2022bkn,Kikuchi:2022lfv,Kikuchi:2022psj} or heterotic string orbifolds~\cite{Baur:2019kwi,Baur:2019iai,Nilles:2020kgo,Nilles:2020tdp,Baur:2020jwc,Nilles:2020gvu,Baur:2020yjl,Baur:2021mtl,Nilles:2021glx,Baur:2021bly,Baur:2022hma,Baur:2024qzo}, where they combine with traditional flavor symmetries, building an eclectic flavor group~\cite{Nilles:2020nnc}. Remarkably, these top-down approaches provide a natural scheme where not only realistic predictions arise~\cite{Baur:2022hma},
but also the modulus $\tau$ can be stabilized close to symmetry-enhanced points~\cite{Font:1990nt,Nilles:1990jv,Cvetic:1991qm,Kachru:2003aw,Intriligator:2006dd,Leedom:2022zdm,Knapp-Perez:2023nty}.

Motivated by these observations, we propose a new supersymmetric model combined with
modular flavor symmetries, which simultaneously accomplishes the following:
\begin{enumerate}
   \item Addressing the flavor puzzle, specifically the origin of the lepton masses and mixing parameters;
   \item Achieving the vacuum alignment for the flavons;
   \item Providing a suitable \ac{DM} candidate with the correct observed \ac{DM} abundance,
         $\Omega_{\text{DM}} = 0.265(7)$~\cite{ParticleDataGroup:2024cfk}.
\end{enumerate}

In order to tackle these issues, we propose a simple supersymmetric model based on a $\Gamma_3\cong A_4$ modular flavor symmetry. The model resembles Model 1 of~\cite{Feruglio:2017spp}, where the neutrino masses arise from a Weinberg operator and the charged-lepton Yukawa couplings are given by the \ac{VEV} of a flavon. It also resembles the proposal of~\cite{Nomura:2024abu}, which studies a \ac{DM} candidate in a non-modular $A_4$ model without fitting flavor parameters. In our model, the flavon potential is fixed by the flavor symmetry together with a $\U1_\mathrm{R}\x\Z2$ symmetry, which determines the couplings between the driving field and flavon superfields. The model gives a very good fit to the leptonic flavor parameters, with a low value of $\chi^{2}$.
Finally, we identify a Dirac fermion composed by the Weyl fermionic parts of both the driving field and flavon superfields; we perform a parameter scan for a correct \ac{DM} abundance.
The goal of this model is to present a ``proof of principle'' that driving fields
in modular supersymmetric flavor models can account for both the flavon \ac{VEV} $\left\langle \phi \right \rangle $ and \ac{DM}.

Our paper is organized as follows. In \Cref{sec:ModularSymmetry}, we review the basics of modular symmetries and its application to solving the flavor puzzle. In \Cref{sec:DarkModularFlavonModel}, we define our model. In \Cref{Sec:FlavorFit}, we present the numerical fit for the lepton flavor parameters. In \Cref{Sec:DarkMatterAbundance}, we analyze the relevant terms for \ac{DM} production. We also argue why we need freeze-in (as opposed to the more traditional freeze-out) mechanism for our model to work. We then present a parameter scan for the available parameter space for our \ac{DM} candidate.
Finally, in \Cref{sec:SummaryAndOutlook} we summarize our results and future directions for further constraints.

\section{Modular symmetry}
\label{sec:ModularSymmetry}

\subsection{Modular groups and modular forms}
\label{subsec:ModularGroupsAndModularForms}

The modular group $\Gamma:=\SL{2,\Z{}}$  is given by
\begin{equation}
\Gamma ~=~ \left\{ \begin{pmatrix} a & b \\ c & d \end{pmatrix} ~\Big|~  a,b,c,d \in \Z{}\; \& \; ad -bc = 1 \right\}\; ,
\label{eq:ModularGroup}
\end{equation}
and can be generated by
\begin{equation}
\label{eq:SAndTGenerator}
S ~=~ \begin{pmatrix}
  0 & 1 \\
  -1 & 0
\end{pmatrix}
\qquad \text{ and } \qquad
T ~=~ \begin{pmatrix}
  1 & 1 \\
  0 & 1
\end{pmatrix}\;,
\end{equation}
which satisfy the general presentation of $\Gamma$, $\langle S,T~|~S^4 = (ST)^3 = \Id, S^2T=T S^2\rangle$.
The principle congruence subgroups of level $N$ of $\Gamma$ are defined as
\begin{equation}
\label{eq:ModularGroupLevelN}
  \Gamma(N) ~:=~ \left\{ \gamma\in \Gamma ~|~ \gamma=\Id \mod N \right\}\,,
\end{equation}
which are infinite normal subgroups of $\Gamma$ with finite index. We can also define the
inhomogeneous modular group $\overline{\Gamma} := \Gamma / \{ \pm \Id \} \cong \mathrm{PSL}(2,\Z{})$ and
its subgroups $\overline\Gamma(N)$ with $\overline\Gamma(2) := \Gamma(2)/\{ \pm \Id \}$ and
$\overline\Gamma(N) := \Gamma(N)$ for $N>2$ (as $-\Id$ does not belong to $\Gamma(N)$).
An element $\gamma$ of a modular group acts on the half-period ratio,  modulus $\tau$, as
\begin{equation}
\label{eq:GammaOnTau}
  \gamma\,\tau ~=~ \frac{a \tau + b}{c\tau + d}\,,\qquad \text{where}\quad \tau\in \mathcal{H}
\end{equation}
and $\mathcal{H}$ is the upper complex half-plane,
\begin{equation}
\label{eq:mathcalH}
  \mathcal{H} ~:=~ \{ \tau \in \mathbbm{C}\; | \; \im \tau > 0 \}\; .
\end{equation}
Modular forms $f(\tau)$ of positive modular weight $k$ and level $N$ are complex functions of $\tau$, holomorphic in $\mathcal{H}$, and transform as
\begin{equation}
\label{eq:ModularFormTrafo}
  f(\tau) ~\stackrel{\gamma}{\longrightarrow}~ f(\gamma\,\tau ) ~=~ (c\tau + d)^k\,f(\tau)\,, \qquad \gamma\in\Gamma(N)\;.
\end{equation}
In this work, we restrict ourselves to even modular weights, $k\in2\mathbbm{N}$, although it is known that modular weights can be odd~\cite{Liu:2019khw} or fractional~\cite{Liu:2020msy,Almumin:2021fbk} in certain scenarios.
Interestingly, modular forms with fixed weight $k$ and level $N$ build finite-dimensional vector spaces,
which close under the action of $\Gamma$. It follows then that they must build representations of a finite
modular group that, for even modular weights, result from the quotient
\begin{equation}
  \Gamma_{N} ~:=~ \overline{\Gamma} / \overline{\Gamma}(N)\,.
\end{equation}
Then, under a finite modular transformation $\gamma \in \Gamma$, modular forms of weight $k$ are $n$-plets
(which are called vector-valued modular forms~\cite{Liu:2021gwa})
$Y(\tau):=\left(f_1(\tau),f_2(\tau),\ldots,f_n(\tau)\right)^\mathrm{T}$ of $\Gamma_{N}$, transforming as
\begin{equation}
\label{eq:TrafoModularForm}
  Y(\tau) ~\stackrel{\gamma}{\longrightarrow}~ Y(\gamma\,\tau ) ~=~ (c\tau+d)^{k}\,\rho(\gamma)\,Y(\tau)\,,
\end{equation}
where $\rho(\gamma)\in \Gamma_{N}$ is a representation of $\gamma$.

\subsubsection[Finite modular group Gamma3 = A4]{\boldmath Finite modular group $\Gamma_3\cong A_4$ \unboldmath}
\label{subsubsec:A4}

As our model is based on $\Gamma_{3} \cong A_{4}$, let us discuss some general features of this group and its modular forms.
$\Gamma_3$ is defined by the presentation
\begin{equation}
\label{eq:presentationOfA4}
  \Gamma_{3} ~=~ \left\langle S,T ~|~ S^2 ~=~(ST)^3 ~=~ T^3 ~=~ \Id \right\rangle \,.
\end{equation}
It has order 12 and the irreducible representations (in the complex basis) are given in \Cref{table:IrrepsA4}.
\begin{table}[b]
	\centering
	\begin{tabular}{ccccc|ccccc}
		\toprule
		          & $\rep{1}$ & $\rep{1}'$ & $\rep{1}''$ & $\rep{3}$ &     & $\rep{1}$ & $\rep{1}'$ & $\rep{1}''$ & $\rep{3}$\\
        \midrule
		$\rho(S)$ & $1$ & $1$              & $1$         & $\frac{1}{3}\begin{pmatrix} -1 & 2 & 2 \\ 2 & -1 & 2 \\ 2 & 2 & -1 \end{pmatrix}$ &
		$\rho(T)$ & $1$ & $\omega$         & $\omega^2$  & $ \begin{pmatrix} 1 & 0 & 0 \\ 0 & \omega & 0 \\ 0 & 0 & \omega^2 \end{pmatrix}$ \\
		\bottomrule
	\end{tabular}
	\caption{Irreducible representations of $\Gamma_3\cong A_4$. Here, $\omega := \e^{\nicefrac{2\pi \I}{3}}$.
	\label{table:IrrepsA4}}
\end{table}
Besides $\rep1^a\otimes\rep1^b=\rep1^c$ with $c=a+b\mod 3$ and $\rep1^a\otimes\rep3=\rep3$, where $a,b,c=0,1,2$ count the number of primes,
we have the nontrivial product rule $\rep{3}\otimes\rep{3}=\rep{1}\oplus \rep{1}'\oplus \rep{1}''\oplus \rep{3}_{S}\oplus \rep{3}_A$,
where $S$ and $A$ stand respectively for symmetric and antisymmetric. Considering two triplets $\rho = ( \rho_1 , \rho_2 , \rho_3 )^\mathrm{T}$
and $\psi = (\psi_1 , \psi_2 , \psi_3 )^\mathrm{T}$, in our conventions the Clebsch-Gordan coefficients of $\rho\otimes\psi$ are
\begin{equation}
\begin{aligned}
  (\rho\otimes\psi )_{\rep{1}} &= \rho_{1} \psi_{1} + \rho_{2} \psi_{3} + \rho_{3} \psi_{2}\,,\quad&
  (\rho\otimes\psi )_{\rep{1}'} &= \rho_{1} \psi_{2} + \rho_{2}\psi_{1} + \rho_{3}\psi_{3}\,,\\
  (\rho\otimes\psi )_{\rep{1}''} &= \rho_{1}\psi_{3} + \rho_{2}\psi_{2} + \rho_{3} \psi_{1}\,, \\
  (\rho\otimes\psi )_{\rep{3}_{S}} &= \frac{1}{\sqrt{3}}\begin{pmatrix}
    2 \rho_{1} \psi_{1} - \rho_{2}\psi_{3} - \rho_{3}\psi_{2} \\
    2 \rho_{3} \psi_{3} - \rho_{1}\psi_{2} - \rho_{2}\psi_{1} \\
    2 \rho_{2} \psi_{2} - \rho_{3}\psi_{1} - \rho_{1}\psi_{3}
  \end{pmatrix},&
  (\rho\otimes\psi )_{\rep{3}_{A}} &= \begin{pmatrix}
    \rho_2 \psi_3 - \rho_3 \psi_2  \\
    \rho_1 \psi_2 - \rho_2 \psi_1  \\
    \rho_3 \psi_1 - \rho_1 \psi_3  \\
  \end{pmatrix}.
\end{aligned}
\end{equation}

The lowest-weight modular forms of $\Gamma_3$ furnish a triplet $Y=(Y_1,Y_2,Y_3)^\mathrm{T}$
of weight $k_Y=2$, whose components are given by~\cite{Feruglio:2017spp}
\begin{equation}
\label{eq:Y3tau}
\begin{aligned}
Y_1(\tau ) & =~ \frac{\I}{2\pi}\left[ \frac{\eta'\left(\frac{\tau }{3}\right)}{\eta\left(\frac{\tau }{3}\right)} + \frac{\eta'\left(\frac{\tau +1 }{3}\right)}{\eta\left(\frac{\tau +1 }{3}\right)} + \frac{\eta'\left(\frac{\tau +2}{3}\right)}{\eta\left(\frac{\tau +2 }{3}\right) }- \frac{27 \eta'(3\tau )}{\eta (3 \tau )}\right] \,, \\
Y_2 (\tau ) & =~  -\frac{\I}{\pi}\left[ \frac{\eta'\left(\frac{\tau }{3}\right)}{\eta\left(\frac{\tau }{3}\right)} + \omega^2\frac{\eta'\left(\frac{\tau +1 }{3}\right)}{\eta\left(\frac{\tau +1 }{3}\right)} + \omega \frac{\eta'\left(\frac{\tau +2}{3}\right)}{\eta\left(\frac{\tau +2 }{3}\right) }\right]\,, \\
Y_3 (\tau ) & =~  -\frac{\I}{\pi}\left[ \frac{\eta'\left(\frac{\tau }{3}\right)}{\eta\left(\frac{\tau }{3}\right)} + \omega\frac{\eta'\left(\frac{\tau +1 }{3}\right)}{\eta\left(\frac{\tau +1 }{3}\right)} + \omega^2 \frac{\eta'\left(\frac{\tau +2}{3}\right)}{\eta\left(\frac{\tau +2 }{3}\right) }\right]\,,
\end{aligned}
\end{equation}
where $\eta( \tau )$ is the so-called Dedekind $\eta$ function
\begin{equation}
\label{eq:DedekindEta}
  \eta ( \tau ) ~=~ q^{\nicefrac{1}{24}}\prod_{n = 1}^{\infty }(1 - q^{n} )
  \qquad\text{with}\qquad q := \e^{2\pi \I \tau }\;.
\end{equation}
Higher-weight modular forms can be constructed
from the tensor products of the weight $2$ modular forms given in \Cref{eq:Y3tau}.

\subsection{Modular supersymmetric theories}
\label{subsec:ModularSupersymmetricTheories}

We consider models with $\mathcal N=1$ {\it global} \ac{SUSY}, defined by the Lagrange density
\begin{equation}
\label{eq:LagrangeDensity}
  \mathcal{L} ~=~ \int \D^2 \theta \,\D^2\bar{\theta}\, K(\Phi, \overline{\Phi}) + \left(\int \D^2\,\theta\, W (\Phi ) + \text{h.c.}\right)\,,
\end{equation}
where $K (\Phi, \overline{\Phi} )$ is the K\"ahler potential, $W (\Phi)$ is the superpotential, and
$\Phi$ denotes collectively all matter superfields $\varphi^i$ of the theory and the modulus $\tau$.

Under an element of the modular symmetry $\gamma\in \Gamma$, $\tau$ transforms according to~\Cref{eq:GammaOnTau}, and matter superfields
are assumed to transform as
\begin{equation}
\label{eq:SuperfieldsTrafos}
  \varphi^i  ~\xyrightarrow{\gamma}~ (c\tau + d)^{-k_i} \rho_{i}(\gamma)\, \varphi^{i} \,,
\end{equation}
where $k_i$ are also called modular weights of the field $\varphi^{i}$, which transform as $\Gamma_N$ multiplets. Modular weights $k_{i}$ are not restricted to be positive integers because
$\varphi^{i}$ are not modular forms. Analogous to \Cref{eq:TrafoModularForm}, the matrix $\rho_{i}(\gamma)$ is a representation of the
finite modular flavor group $\Gamma_N$.

For simplicity, we assume a minimal K\"ahler potential\footnote{In principle, there could be further
terms in the K\"ahler potential with an impact on the flavor predictions~\cite{Chen:2019ewa}, which are ignored here.}
of the form
\begin{equation}
\label{eq:MinimalKahler}
  K (\Phi, \overline{\Phi} ) ~=~ -\log( -\I \tau + \I \bar{\tau } ) + \sum_{i}(-\I \tau + \I \bar{\tau} )^{-k_i} |\varphi^{i} |^2 \,.
\end{equation}
Making use of \Cref{eq:SuperfieldsTrafos}, we see that $K(\Phi, \overline{\Phi})$ transforms
under a modular transformation $\gamma\in\Gamma$ as
\begin{equation}
\label{eq:KTrafo}
 K (\Phi, \overline{\Phi} ) ~\xyrightarrow{\gamma }~  K (\Phi, \overline{\Phi} ) + \log (c\tau +d ) + \log (c\bar{\tau} +d )\,.
\end{equation}
Thus, realizing that the K\"ahler potential is left invariant up to a {\it global} supersymmetric K\"ahler transformation,
in order for the Lagrange density of \Cref{eq:LagrangeDensity} to be modular invariant, we need the superpotential
to be invariant under modular transformations, i.e.
\begin{equation}
\label{eq:ModularTrafoOfSuperpotential}
  W (\Phi ) ~\xyrightarrow{\gamma }~ W (\Phi ) \,.
\end{equation}
The superpotential $W (\Phi)$ has the general form
\begin{equation}
W (\Phi ) ~=~ \mu_{ij}(\tau )\varphi^{i}\varphi^{j} + Y_{ijk}(\tau)\varphi^{i}\varphi^{j}\varphi^{k} + G_{ijk\ell }(\tau)\varphi^{i}\varphi^{j}\varphi^{k}\varphi^{\ell } \; ,
\label{eq:Superpotential}
\end{equation}
where $\mu_{ij}(\tau )$, $Y_{ijk}(\tau )$ and $G_{ijk}(\tau )$ are modular forms of level $N$.
Because of \Cref{eq:ModularTrafoOfSuperpotential}, each term of \Cref{eq:Superpotential} must be modular invariant.
Let us illustrate how we can achieve this by taking the trilinear coupling
$Y_{ijk}(\tau)\varphi^{i}\varphi^{j}\varphi^{k}$. The Yukawa coupling $Y_{ijk}$ transforms
under a modular transformation $\gamma\in\Gamma$ as
\begin{equation}
\label{eq:YukawacouplingTrafo}
  Y_{ijk}(\tau ) ~\xyrightarrow{\gamma}~ (c\tau + d)^{k_Y}\rho_{Y} (\gamma)\,Y_{ijk}(\tau ) \,,
\end{equation}
where $k_Y$ is the even integer modular weight of the modular form $Y_{ijk}(\tau)$. Then, for $Y_{ijk}(\tau)\varphi^{i}\varphi^{j}\varphi^{k}$
to be invariant and using the superfield transformations of \Cref{eq:SuperfieldsTrafos},
we must demand that $k_{Y} = k_{i}+k_j + k_k $ and that the product
$\rho_{Y} \otimes \rho_{i} \otimes \rho_{j} \otimes \rho_{k}$ contains an invariant singlet.

Since we shall be concerned with \ac{SUSY} breaking, let us briefly discuss the soft-\ac{SUSY} breaking terms in the Lagrange density.
They are given by
\begin{equation}
\label{eq:SoftTerms}
  \mathcal{L}_{\text{soft}} ~=~ -\frac{1}{2} \left( M_{a} \hat{\lambda}^{a}\hat{\lambda}^{a} + \text{h.c} \right)
                              - \tilde{m}_{i}^{2}\bar{\hat{\varphi}}^{i}\hat{\varphi}^{i}
                              - \left(A_{ijk}\hat{\varphi}^{i}\hat{\varphi}^{j}\hat{\varphi}^{k} + B_{ij}\hat{\varphi}^{i}\hat{\varphi}^{j} + \text{h.c.}\right)\; ,
\end{equation}
where $M_{a}$ are the gaugino masses, $\hat{\lambda}^{a}$ are the canonically normalized gaugino fields, and $\tilde{m}_{i}$ are the soft-masses.
We use a notation, where $\hat{\varphi}^{i}$ stands for both the canonically normalized chiral superfield and its scalar component~\cite{Ding:2022nzn}. We do not assume any specific source of \ac{SUSY} breaking, and thus the parameters in \Cref{eq:SoftTerms} are free, in principle. However, in our model $B-$terms are forbidden at tree level and the $A$-terms do not play a role in DM production at tree-level, see \Cref{Sec:DarkMatterAbundance}.

\section{Dark modular flavon model}
\label{sec:DarkModularFlavonModel}

We propose a model of leptons, governed by the modular flavor group
$\Gamma_{3} \cong A_{4}$. In our model, the lepton doublets $L$ transform as a flavor triplet and charged-lepton singlets $E^\mathrm{c}_i$ transform as three distinct singlets under $\Gamma_3$. The particle content of our model is summarized in \Cref{table:Model}. 
\begin{table}[h!]
	\centering
	\begin{tabular}{lccccccccc}
		\toprule
		& $L$     & $(E_1^\mathrm{c}, E_2^\mathrm{c}, E_3^\mathrm{c})$   & $H_d$   & $H_u$   & $\phi_3$ & $\phi_{1'}$ &  $\zeta_3$& $\zeta_{1''}$ & $Y(\tau )$ \\
		\midrule
		$\SU2_{L}$   & $\rep2$ & $\rep1$ & $\rep2$ & $\rep2$ & $\rep1$  & $\rep1$   & $\rep1$    & $\rep1$     & $\rep1 $    \\
		\arrayrulecolor{lightgray}\midrule\arrayrulecolor{black}
		$\U1_{Y}$   & $-\frac{1}{2}$ & $1$ & $-\frac{1}{2}$ & $\frac{1}{2}$ & $ 0$  & $ 0$   & $ 0$    & $0$     & $0$    \\
		\arrayrulecolor{lightgray}\midrule\arrayrulecolor{black}
		$\Gamma_3\cong A_4$   & $\rep3$ & $(\rep1,\rep1'', \rep1')$ & $\rep1$ & $\rep1$ & $\rep3$  &  $\rep1'$   &   $\rep3$ & $\rep1''$     & $\rep3$    \\
		\arrayrulecolor{lightgray}\midrule\arrayrulecolor{black}
		$k_i$ & $1$ & $0$                       & $-1$    & $0$     & $0$      & $0$       & $0$         & $0$           & $2$        \\
		\arrayrulecolor{lightgray}\midrule\arrayrulecolor{black}
		$\U1_\mathrm{R}$ & $1$   & $1$                & $0$     & $0$     & $0$      & $0$       & $2$         & $2$           & $0$        \\
		\arrayrulecolor{lightgray}\midrule\arrayrulecolor{black}
		$\Z2$     & $0$     & $0$                       & $0$     & $0$     & $0$      & $-1$       & $0$ & $-1$ & $0$        \\
		\bottomrule
	\end{tabular}
	\caption{Quantum numbers under the SM gauge groups, $\U1_\mathrm{R}$, $\Z2$ and $\Gamma_3$, as well as modular weights
	         $k_i$ of the matter fields in our model. All fields are neutral under $\SU3_{\mathrm{color}}$.
	         \label{table:Model}}
\end{table}

In our model, neutrino masses arise from the Weinberg operator
\begin{align}
\label{eq:Wnu_model}
    W_{\nu} ~=~ \frac{1}{\Lambda}\left( H_u L H_u L Y(\tau ) \right)_{\rep{1}}\;,
\end{align}
where $\Lambda$ is the neutrino-mass scale and $Y(\tau)$ is the modular-form
triplet $Y=(Y_1,Y_2,Y_3)^\mathrm{T}$ of weight $2$ given by \Cref{eq:Y3tau}. 
Note that there exists no other modular multiplet of weight $2$ in $\Gamma_3\cong A_4$. Hence, under our assumptions, the neutrino sector obtained from \Cref{eq:Wnu_model} is highly predictive as it only depends on the parameter $\tau$, the \ac{VEV} $v_u$ of $H_u$, and $\Lambda$.

The charged-lepton superpotential at leading order is
\begin{align}
    W_{\mathrm{CL}}
    ~=~ \frac{\alpha_1}{\Lambda_\phi} E_1^\mathrm{c} H_d (L \phi_{3})_{\rep{1}}+ \frac{\alpha_2}{\Lambda_\phi} E_2^\mathrm{c} H_d (L \phi_{3})_{\rep{1}'}+\frac{\alpha_3}{\Lambda_\phi} E_3^\mathrm{c} H_d (L \phi_{3})_{\rep{1}''}\;,
    \label{eq:W_CL_model}
\end{align}
where $\alpha_{i}$, $i=1,2,3$, are dimensionless parameters, $\Lambda_{\phi}$ denotes the flavor breaking scale, and the subindices refer to the respective $\Gamma_3$ singlet components of tensored matter states in the parentheses. The charged-lepton mass matrix can be determined by the $\Gamma_3$ triplet flavon \ac{VEV} $\left\langle \phi_{3} \right\rangle$ as in Model 1 of~\cite{Feruglio:2017spp}. However, we have taken a different value of $\left\langle \phi_{3} \right\rangle $, which eventually leads to a better fit.

The flavon superpotential is given by
\begin{align}
    W_{\phi} &= \Lambda_\phi \beta_1  \zeta_3 \phi_3 + \beta_2\zeta_3 \phi_3 \phi_3  \; +  \frac{\beta_3}{\Lambda_\phi}\zeta_3 \phi_3 \phi_3 \phi_3 + \Lambda_\phi \beta_4 \zeta_{1''}\phi_{1'} + \frac{\beta_5}{\Lambda_\phi} \zeta_{1''}\phi_{1'}\phi_3\phi_3 \nonumber \\
    & + \frac{\beta_6}{\Lambda_\phi} \zeta_3 \phi_3 \phi_{1'} \phi_{1'}\;,
    \label{eq:Wphi_model}
\end{align}
where $\beta_{i}$, $i=1,\ldots ,6$, are dimensionless couplings. This superpotential gives rise to the desired \ac{VEV} pattern with driving superfields $\zeta_{r}$, $r = 1'', 3 $, and an extra flavon $\phi_{1'}$, where the subindices label the respective $A_4$ representations. To fix the flavon superpotential, we impose a symmetry $\U1_{\mathrm{R}} \x\Z2$, similarly to~\cite{Altarelli:2005yx,Nomura:2024abu}. As usual, the $\U1_{\mathrm{R}}$ $R$-symmetry forbids the renormalizable terms in the superpotential that violate lepton and/or baryon numbers.
We remark that the flavon superpotential \Cref{eq:Wphi_model} is important for both finding the correct vacuum alignment for the flavons, as discussed below, and for identifying a viable \ac{DM} candidate as described in \Cref{Sec:DarkMatterAbundance}.

The flavon \ac{VEV} is then attained by demanding that \ac{SUSY} remains unbroken at a first stage, i.e.~we require vanishing $F$-terms.
Recall that the $F$-term scalar potential in a global supersymmetric theory is given schematically by
\begin{equation}
 V ~=~ F^{i}K_{i\bar{\jmath}}\bar{F}^{\bar{\jmath}}\; ,
\label{eq:ScalarPotentialSupergravity}
\end{equation}
where
\begin{equation}
F^{i}  = -\frac{\partial W}{\partial \Phi^{i}}\,,
\qquad
\bar{F}^{\bar{\jmath}} = F^{j*} 
\qquad\text{and}\qquad
K^{i\bar{\jmath}} = \left( K_{i\bar{\jmath}}\right)^{-1} = (-\I \tau + \I \tau)^{k_i}\delta_{i\bar{j}}\; .
\label{eq:CovariantFi}
\end{equation}

Recall that a superfield $\Phi$ can be expanded in its components as~\cite[Equation 2.117]{Terning:2006bq}
\begin{align}
\Phi &= \Phi (x) - \I\theta\sigma^{\mu}\bar{\theta}\partial_\mu \Phi (x) - \frac{1}{4}\theta^2 \bar{\theta}^2\partial^2 \Phi  (x) + \sqrt{2}\theta\psi_{\Phi} (x)  \nonumber\\
&+ \frac{\I}{\sqrt{2}}\theta^2\partial_\mu \psi_{\Phi} (x)\sigma^{\mu}\bar{\theta}+\theta^2 F_{\Phi} (x)\;,
\label{eq:SuperfieldComponentsVarphi}
\end{align}
where we have used the notation that $\Phi$ represents both the superfield and its scalar component, $\psi_{\Phi}$ is the fermionic component and $F_{\Phi}$ is the $F$-term.
For \ac{SUSY} to be preserved, we must have $\left\langle F_{\Phi} \right\rangle = 0$. We assume that the only possible sources of \ac{SUSY} breaking are given either by $\tau $ or a hidden sector. Thus, we demand $\left\langle F_{\varphi^i} \right\rangle = 0$, for all $i$, where $\varphi^i$ represents the matter fields in our model (cf.~\Cref{eq:LagrangeDensity}).
We solve these $F$-term equations at the \ac{VEV}'s of the flavons $\phi_3$ and $\phi_{1'}$ and Higgs fields,
\begin{equation}
\left\langle \phi_3 \right\rangle  =: v_3 \left( 1, a, b\right)^{T}, \qquad \left\langle \phi_{1'} \right\rangle  =: v_{1'} \;, \qquad
\left\langle H_u \right\rangle  =: v_u \qquad\text{and}\qquad \left\langle H_d \right\rangle  =: v_d \;.
\label{eq:FlavonVevs}
\end{equation}
where we assume $a,b,v_3,v_{1'} \in \mathbbm{R}$. All $F$-term equations are trivially satisfied except the ones corresponding to the driving fields given by 
\begin{equation}
\label{eq:DrivingFieldsFterm}
  \langle F_{\zeta_{3i}}\rangle ~=~ 0 \qquad\text{and}\qquad
  \langle F_{\zeta_{1''}}\rangle ~=~ 0\;,
\end{equation}
where $\zeta_{3i}$, $i=1,2,3$, are the three components of the triplet. Thus, we obtain the following relations
\begin{equation}
\label{eq:couplings_Ftermab}
\begin{aligned}
    \beta_2 &~=~ c_2(a,b)  \beta_1 \frac{ \Lambda_\phi }{v_3}\,,   \qquad& \beta_3 &~=~ c_3(a,b) \beta_1\frac{ \Lambda_\phi^2 }{v_3^2}\;,\\
	\beta_5 &~=~ c_5(a,b) \beta_4\frac{ \Lambda_\phi^2 }{v_3^2}\,, & \beta_6 &~=~ c_6(a,b) \beta_1\frac{ \Lambda_\phi^2 }{v_{1'}^2}\;,
\end{aligned}
\end{equation}
where $c_i(a,b)$, for $i=2,3,5,6$, are coefficients that depend only on the values of $a,b$. The numerical values of $a,b$ dictate the charged lepton mass matrix. Hence, they are determined by the fit to the flavor parameters that we do in \Cref{Sec:FlavorFit}. Through the flavor parameter fit, we have found values of $\left\langle \phi_3\right\rangle, \left\langle \phi_{1'} \right \rangle$ that satisfy simultaneously \Cref{eq:couplings_Ftermab}, thus yielding vacuum alignment.

Finally, we assume that the flavon \ac{VEV} scale from \Cref{eq:FlavonVevs} is below $\Lambda_{\phi}$, such that
\begin{equation}
\label{eq:FlavonVevsScale}
  v_3  ~=~ v_{1'}  ~=~ 0.1 \Lambda_{\phi}\;.
\end{equation}
Furthermore, we identify the \ac{DM} candidate as a Dirac fermion built as a combination of the Weyl components of $\zeta_{i}$ and $\phi_{i}$ with the scalar component of the flavon $\phi_{i}$ serving as a mediator.
The parameters in \Cref{eq:FlavonVevsScale} shall play an important role in finding the current \ac{DM} abundance since they set the couplings in \Cref{eq:Wphi_model}.

\section{Flavor fit}
\label{Sec:FlavorFit}

Having defined our model of modular flavored dark matter, we next assess its capability to reproduce the experimentally observed charged-lepton masses, neutrino squared mass differences, and the mixing parameters of the PMNS matrix
while providing predictions for yet undetermined observables, such as the three absolute neutrino masses, the Dirac \CP phase, and the two Majorana phases.

The explicit neutrino mass matrix can be determined from \Cref{eq:Wnu_model}. By calculating the tensor products to obtain the symmetry invariant part,
we find that the neutrino mass matrix is predicted to be
\begin{equation}
    M_\nu ~=~ \frac{v_u^2}{\Lambda} \begin{pmatrix} 2 Y_1 & -Y_3 & -Y_2 \\ -Y_3 & 2 Y_2 & -Y_1 \\ -Y_2 & -Y_1 & 2 Y_3 \end{pmatrix}.
    \label{eq:Ynu_mass_modular_Dirac}
\end{equation}
The charged-lepton mass matrix $M_{\text{CL}}$ arises from \Cref{eq:W_CL_model}. Substituting the flavon \ac{VEV}s as defined in \Cref{eq:FlavonVevs} and calculating the tensor products,
we arrive at
\begin{equation}
    M_{\text{CL}} ~=~ \frac{v_d v_3}{\Lambda_\phi}\begin{pmatrix}
        \alpha_1  & \alpha_1 b & \alpha_1 a\\
         \alpha_2 a  & \alpha_2   & \alpha_2 b \\
        \alpha_3 b  &  \alpha_3 a & \alpha_3
    \end{pmatrix} ~=~ 0.1 v_d\begin{pmatrix}
        \alpha_1  & \alpha_1 b & \alpha_1 a\\
         \alpha_2 a  & \alpha_2   & \alpha_2 b \\
        \alpha_3 b  &  \alpha_3 a & \alpha_3
    \end{pmatrix}\; ,
    \label{eq:Mcl_model}
\end{equation}
where we have used \Cref{eq:FlavonVevsScale}. The values of $v_u$ and $v_d$ are determined by the Higgs \ac{VEV}, $v = \sqrt{v_u^2 + v_d^2} =246$\,GeV, and $\tan\beta = \frac{v_u}{v_d}$, which we assume to be $\tan \beta = 60 $. The neutrino mass scale is determined by $\Lambda$, while we choose $\alpha_3$ to set the mass scale of charged leptons. By using the standard procedure (see e.g.\ ref.~\cite{ParticleDataGroup:2024cfk}), one arrives at the lepton masses and the PMNS mixing matrix.
For our model, the resulting 12 flavor observables depend on 6 real dimensionless parameters $\re\tau$, $\im\tau$, $a$, $b$, $\alpha_1/\alpha_3$, and $\alpha_2/\alpha_3$, as well as two dimensionful overall mass scales $v_u^2/\Lambda$ and $0.1 v_d \alpha_3$.

\begin{table}[t]
	\centering
	\begin{tabular}{ll}
		\toprule
		observables                                  & $\phantom{-}$best-fit values  \\
		\midrule
		$m_\mathrm{e}/m_\mu$                         & $\phantom{-}0.00473\pm0.00004$ \\
		$m_\mu/m_\tau$                               & $\phantom{-}0.0450 \pm 0.0007$ \\
		$y_\tau$                                     & $\phantom{-}0.795 \pm 0.012$ \\
		\arrayrulecolor{lightgray}\midrule\arrayrulecolor{black}
		$\Delta m_{21}^2 / 10^{-5} ~[\mathrm{eV}^2]$ & $\phantom{-}7.41^{+0.21}_{-0.20}$\\[4pt]
		$\Delta m_{32}^2 / 10^{-3} ~[\mathrm{eV}^2]$ & $-2.487^{+0.027}_{-0.024}$\\[4pt]
		$\sin^2\theta_{12}$                          & $\phantom{-}0.307^{+0.012}_{-0.011}$ \\[4pt]
		$\sin^2\theta_{13}$                          & $\phantom{-}0.02222^{+0.00069}_{-0.00057}$\\[4pt]
		$\sin^2\theta_{23}$                          & $\phantom{-}0.568^{+0.016}_{-0.021}$\\[4pt]
		$\delta_{\CP}^{\ell}/\pi$                    & $\phantom{-}1.52^{+0.13}_{-0.15}$\\[2pt]
		\bottomrule
	\end{tabular}
	\caption{
		Experimental central values and $1\sigma$ uncertainties for the masses and mixing parameters of the lepton sector.
		The data for the neutrino oscillation parameters is taken from the global analysis NuFIT v5.3~\cite{Esteban:2020cvm}
		for inverted ordering taking the Super-Kamiokande data into account.
		The charged-lepton mass ratios and the tau Yukawa coupling $y_\tau$ are taken from ref.~\cite{Antusch:2013jca} with $M_\mathrm{SUSY}=10\,\mathrm{TeV}$, $\tan\beta = 60$, and $\bar\eta_b=0$.
    	}
		\label{tab:ExpDataLeptons}
\end{table}

To show that the model can accommodate the observed flavor structure of the \ac{SM} lepton sector,
we scan its parameter space and compare the resulting flavor observables to experimental data,
with the best-fit values shown in \Cref{tab:ExpDataLeptons}.
As an approximate measurement of the goodness of our fit, we introduce a $\chi^2$ function
\begin{equation}
\chi^2 ~=~ \sum_{i} \chi_i^2\,,
\end{equation}
consisting of a quadratic sum of one-dimensional chi-square projections for each observable.
Here, we assume that the uncertainties of the fitted observables are independent of each other
and do not account for the small correlations among experimental errors of $\sin^2\theta_{23}$ and other quantities.
For the mixing angles and the neutrino squared mass differences, we determine the value of $\chi_i^2$
directly from the one-dimensional projections determined by the global analysis NuFIT v5.3~\cite{Esteban:2020cvm},
which are available on their website. This is necessary
to account for the strong non-Gaussianities in the uncertainties of the mixing parameters in the PMNS matrix.
For these observables, we refrain from considering corrections from renormalization group running,
given that their contribution is expected to be small compared to the size of the experimental errors.
For the charged-lepton masses, we determine the value of $\chi_i^2$ by
\begin{equation}
\chi_i ~=~ \dfrac{\mu_{i,\mathrm{exp.}} - \mu_{i,\mathrm{model}}}{\sigma_i}\;,
\end{equation}
where $\mu_{i,\mathrm{model}}$ denotes the resulting value for the $i$th observable of the model,
while $\mu_{i,\mathrm{exp.}}$ and $\sigma_i$ refer to its experimentally observed central value
and the size of the $1\sigma$ uncertainty interval given in \Cref{tab:ExpDataLeptons}, respectively.
The total value of $\chi^2$ for all considered observables may then be interpreted to indicate an
agreement with the experimental data at a $\sqrt{\chi^2}\,\sigma$ confidence level (C.L.).

\begin{figure}[t!]
	\centering
	\includegraphics[width=\textwidth]{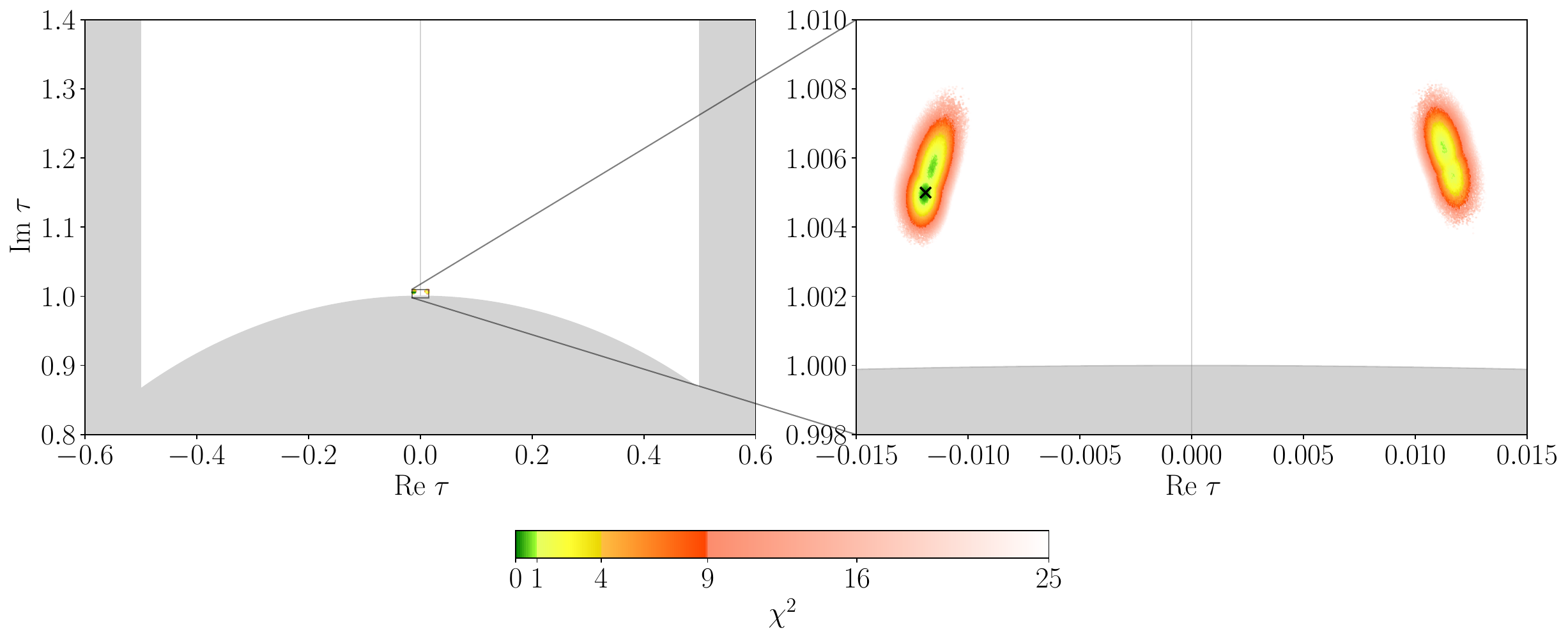}
	\caption{Regions in moduli space that yield fits with $\chi^2\leq 25$. The green, yellow, and orange colored regions may be interpreted as the $1\sigma$, $2\sigma$, and $3\sigma$ confidence intervals, respectively, while the opaque red fades out until the $5\sigma$ barrier is reached. The best-fit point of \Cref{eq:BestFitFlavor} is marked by $\mathsf{x}$. The unshaded area corresponds to the fundamental domain of $\SL{2,\mathbbm{Z}}$.
}
	\label{fig:Modulispace}
\end{figure}

\vspace{3.5mm}
To scan the parameter space of the model and minimize the $\chi^2$ function, we use the dedicated code
\texttt{FlavorPy}~\cite{FlavorPy}. We find that the model is
in agreement with current experimental observations.
The best-fit point in the parameter space of our model is at
\begin{align}
\tau &= -0.0119 +   1.005 \I\;,        & a&=-0.392\;, & b&=0.380\;, & 0.1 \alpha_3 v_d & = 0.130\,\mathrm{GeV} \nonumber\\
\frac{\alpha_1}{\alpha_3} &= -22.0\;, &\frac{\alpha_2}{\alpha_3}&=4.78\x 10^{-3}\;, &\frac{v_u^2}{\Lambda} &=0.0221\,\mathrm{eV}\;,
\label{eq:BestFitFlavor}
\end{align}
where we obtain $\chi^2=0.08$, meaning that all resulting observables are within their experimental $1\sigma$ interval, cf.\ also \Cref{eq:BestFitFlavorObservables}.
In \Cref{fig:Modulispace}, we present the regions in moduli space that yield results with $\chi^2\leq 25$.
\vspace{3.5mm}

\begin{figure}[h!]
	\centering
	\includegraphics[width=\textwidth]{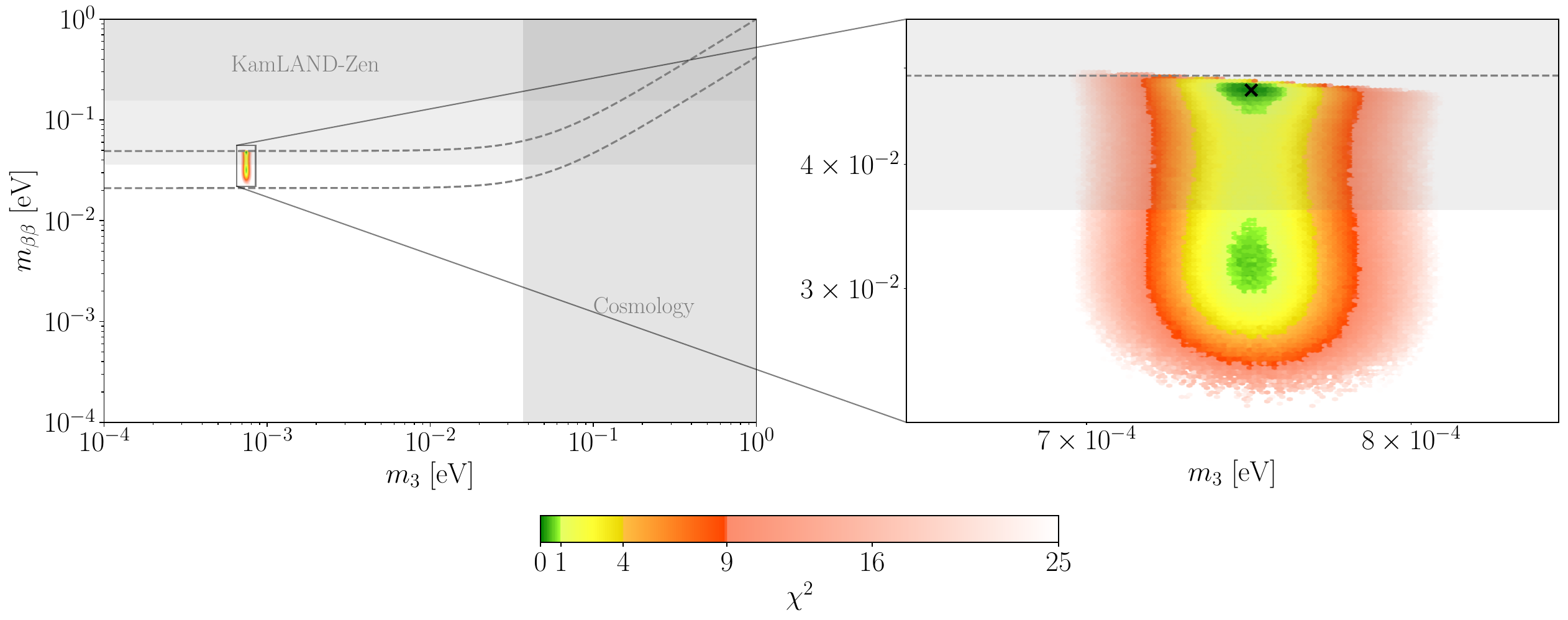}
	\caption{The effective neutrino mass for neutrinoless double beta decay as a function of the lightest neutrino mass. The experimentally allowed region within $3\sigma$ for inverted ordering is the region between the two dashed lines, while the region allowed in our model is represented by the colored area. The gray-shaded areas are excluded by KamLAND-Zen~\cite{KamLAND-Zen:2022tow} or cosmological bounds~\cite{GAMBITCosmologyWorkgroup:2020rmf,Planck:2018vyg}. We remark that the KamLAND-Zen upper limit for the effective neutrino mass $m_{\beta\beta}$ depends on the used nuclear matrix element estimate.
	While all estimates applied by the KamLAND-Zen collaboration give rise to an upper bound of the effective mass to be below $156\,$meV (dark gray area), only some estimates yield limits that are smaller than $36\,$meV (light gray area)~\cite{KamLAND-Zen:2022tow}.
	Hence, we find that the sample point marked by $\mathsf{x}$ is only challenged by some nuclear matrix element estimates while still being allowed according to the majority of estimates.
	}
	\label{fig:m3mbb}
\end{figure}

For the specific values given in \Cref{eq:BestFitFlavor}, the relations among the couplings of the flavon superpotential of \Cref{eq:couplings_Ftermab} read
\begin{align}
\beta_2 = 4.27  \beta_1\; , \qquad
\beta_3 = 83.5 \beta_1\;, \qquad
\beta_5 = 142 \beta_4\;, \qquad
\beta_6 = 121 \beta_1\,.
\label{eq:couplings_FtermNumerical}
\end{align}
Any values of $\beta_1$ and $\beta_4$ then solve the $F$-term equations of the driving fields, cf.\ \Cref{eq:DrivingFieldsFterm}, and ensure the specific vacuum alignment of \Cref{eq:BestFitFlavor}. 
The resulting observables at the best-fit point given in \Cref{eq:BestFitFlavor} lie well within the $1\sigma$ intervals of the experimental data shown in \Cref{tab:ExpDataLeptons}, and read
\begin{equation}
\begin{aligned}
m_{\mathrm{e}}/m_\mu &= 0.00473\;, & m_\mu/m_\tau &= 0.0451\;,\quad& y_\tau & = 0.795\; , \\
\sin^2\theta_{12} &= 0.306\;,           & \sin^2\theta_{13} &= 0.02231\;, \quad& \sin^2\theta_{23} &= 0.568\;,\\
\delta_{\CP}^\ell/\pi &= 1.52 \; \text{rad} \, , & \eta_1/\pi &= 1.41  \; \text{rad}, & \eta_2/\pi &= 0.351  \; \text{rad} \, ,\\
m_1 &= 49\,\mathrm{meV}\;, & m_2 &= 50\,\mathrm{meV}\;, & m_3 &= 0.75\,\mathrm{meV}\;.
\end{aligned}
\label{eq:BestFitFlavorObservables}
\end{equation}
Moreover, the resulting sum of neutrino masses, the neutrino mass observable in $^3\mathrm{H}$ beta decay, and the effective neutrino mass for neutrinoless double beta decay, are
\begin{equation}
\sum m_i = 100\,\mathrm{meV} \;, \qquad
m_\beta = 50\,\mathrm{meV}\,, \qquad\text{and}\qquad
m_{\beta\beta} = 48\,\mathrm{meV}\,,
\end{equation}
which are consistent with their latest experimental bounds
$\sum m_i < 120\,\mathrm{meV}$~\cite{Planck:2018vyg}, $m_\beta < 800\,\mathrm{meV}$~\cite{KATRIN:2021uub}, and $m_{\beta\beta} < 156\,\mathrm{meV}$~\cite{KamLAND-Zen:2022tow}.
It is to be noted that our predicted value of the effective neutrino mass $m_{\beta\beta}$ is challenged by experimental bounds determined with certain nuclear matrix element estimates \cite{KamLAND-Zen:2022tow}, as illustrated in \Cref{fig:m3mbb}.

\begin{figure}[h!]
	\centering
	\includegraphics[width=0.55\textwidth]{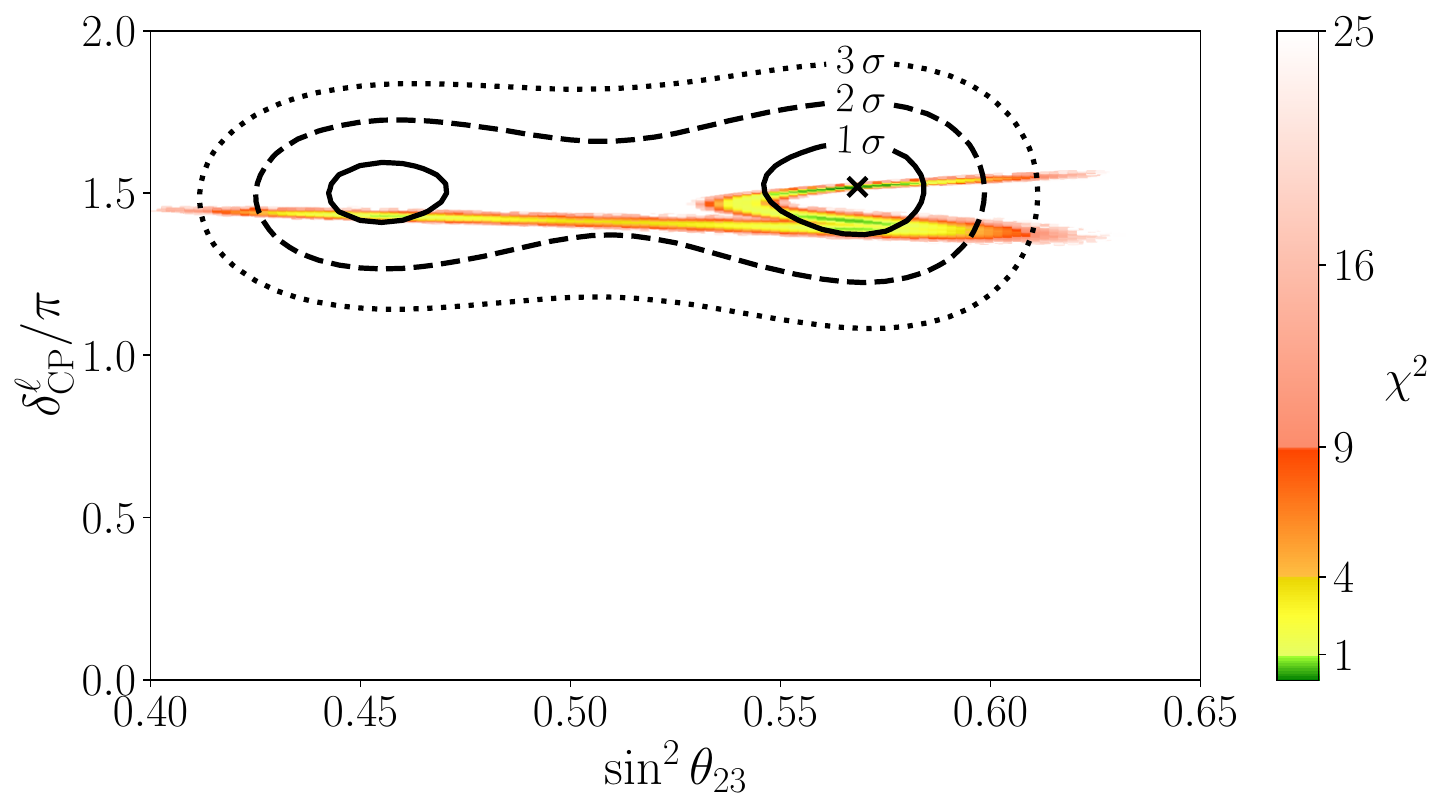}
	\caption{Predicted regions in the space of $\sin^2 \theta_{23}$ and $\delta_{\CP}^\ell$ with $\chi^2 \leq 25$ in our model.
	The black lines indicate the regions that are experimentally admissible at a $1$, $2$, and $3\sigma$ C.L.\ as obtained by the global analysis NuFIT v5.3~\cite{Esteban:2020cvm}.
	\label{fig:s23dcp}}
\end{figure}

We remark that the model can be consistent with both octants of $\theta_{23}$, while only being compatible with Dirac \CP-violating phases in the range of $1.36 < \delta_{\CP}^\ell < 1.55$ at a $3\sigma$ C.L., as shown in \Cref{fig:s23dcp}.
Moreover, the inverted-ordering neutrino masses are predicted to lie within the narrow ranges
\begin{equation}
48\,\mathrm{meV} < m_1 < 50\,\mathrm{meV}\;, \quad
49\,\mathrm{meV} < m_2 < 51\,\mathrm{meV}\;, \quad
0.72\,\mathrm{meV} < m_3 < 0.78\,\mathrm{meV}\;,
\end{equation}
at a $3\sigma$ C.L. The numerical analysis suggests that the model prefers a neutrino spectrum with inverted ordering.
For a normal-ordering spectrum we only obtain a match with experimental data just barely in the $3\sigma$ interval with $\chi^2\approx 7$.

\newpage

\section{DM abundance}
\label{Sec:DarkMatterAbundance}

Let us start by identifying the \ac{DM} candidate. Since the flavon field only couples to charged leptons,
the phenomenological implications for \ac{DM} in our model are determined by the charged-lepton Yukawa interactions and the flavon potential.
The interactions between the \ac{SM} charged leptons and the flavon scalar $\phi_{3}$ are given by
\begin{align}
    \mathcal{L}_{\mathrm{CL}} & ~=~ -\sum_{i=1}^{3}\frac{v_d}{2} \alpha_i\, \psi_{E_{i}^{C}} \left(\psi_{E}\phi_{3}\right)_{\rep{1}} +\text{h.c} \;       \label{eq:L_CL}\\
   & ~\longrightarrow~ -\sum_{i=1}^{3}\frac{v_d}{2} \alpha_i \, \psi_{E_{i}^{C}} \left(\psi_{E}     \left(\phi_{3}  +v_3 (1,a,b)^{T}\right)\right)_{\rep{1}} +\text{h.c.}\;,\nonumber
\end{align}
where $\psi_E$ denotes the fermionic part of the charged-lepton component of $L$. In the last line, we develop the flavon $\phi_{3}$ in the vacuum with its \ac{VEV} given by $v_3 (1,a,b)^{T}$, see \Cref{eq:FlavonVevs}. From the leading flavon superpotential terms in \Cref{eq:Wphi_model},\footnote{We ignore the terms suppressed by $\Lambda_{\phi}$.} we obtain
\begin{align}
    \mathcal{L}_\phi &~=~ -\frac{1}{2} \Lambda_\phi \beta_1 \left( \psi_{\zeta_{3}} \psi_{\phi_{3}}\right)_{\rep{1}} -\frac{1}{2} \beta_2 \left(\psi_{\zeta_{3}} \psi_{\phi_{3}} \phi_{3}\right)_{\rep{1}} -\frac{1}{2} \Lambda_\phi \beta_4 \left(\psi_{\zeta_{1''}} \psi_{\phi_{1'}}\right)_{\rep{1}}+\text{h.c.}   \label{eq:L_phi}\\
    &~\to~ -\frac{1}{2} \Lambda_\phi \beta_1 \left( \psi_{\zeta_{3}} \psi_{\phi_{3}}\right)_{\rep{1}} -\frac{1}{2} \beta_2 \left(\psi_{\zeta_{3}} \psi_{\phi_{3}} \left(\phi_{3}  +v_3 (1,a,b)^{T}\right)\right)_{\rep{1}} \nonumber \\
    & ~~~~~ -\frac{1}{2} \Lambda_\phi \beta_4 \left(\psi_{\zeta_{1''}} \psi_{\phi_{1'}}\right)_{\rep{1}}+\text{h.c.} \nonumber\\
    &~=~ -\frac{1}{2} \Lambda_\phi \beta_1 \left( \psi_{\zeta_{3}} \psi_{\phi_{3}}\right)_{\rep{1}} -\frac{4.27}{2} \beta_1 \left(\psi_{\zeta_{3}} \psi_{\phi_{3}} \left(\phi_{3}  +v_3 (1,a,b)^{T}\right)\right)_{\rep{1}}\nonumber\\
    &~~~~~ -\frac{1}{2} \Lambda_\phi \beta_4 \left(\psi_{\zeta_{1''}} \psi_{\phi_{1'}}\right)_{\rep{1}}+\text{h.c.}\,,\nonumber
\end{align}
where $\phi_3$ is again expanded around its \ac{VEV} in the second line and we use the relations of \Cref{eq:couplings_FtermNumerical} in the last line.

The \ac{DM} candidate is the lowest mass-state Dirac fermion built as a linear combination of the Weyl components of the driving fields and flavon fields, $\psi_{\zeta_i}, \psi_{\phi_i}$, whereas the mediator is a linear combination of the flavon scalars. The particle content relevant for \ac{DM} production is outlined in \Cref{table:ModelDM}. After the Higgs and the flavons acquire \ac{VEV}s as given in \Cref{eq:FlavonVevs},\footnote{The reheating temperature is chosen to be $T_{\text{R}} =150$\,GeV such that \ac{DM} production begins after \ac{EW} symmetry breaking. Recall that the \ac{EW} symmetry breaking temperature of crossover is around $160$\,GeV with a width of $5$\,GeV~\cite{DOnofrio:2015gop}. We acknowledge that higher reheating temperatures are also possible, in which case, the production of \ac{DM} happens before \ac{EW} symmetry breaking. We leave this possibility for future work.} the symmetry group of \Cref{table:Model} gets broken down to $\U1_{\text{EM}}\times \U1_{R}$.

\begin{table}[t]
	\centering
	\begin{tabular}{lcccccccc}
		\toprule
		& $\psi_{E_{i}}$     & $\psi_{E_{i}^C}$      & $\phi_{3,i}$ & $\phi_{1'}$ & $\psi_{\phi_{3,i}}$ & $\psi_{\zeta_{3,i}}$ & $\psi_{\phi_{1'}}$ & $\psi_{\zeta_{1''}}$ \\
		\midrule
		$\U1_{\text{EM}}$   & $-1$ & $+1$ & $0$ & $0$ & $0$ & $0$ & $0$ & $0$    \\
		\arrayrulecolor{lightgray}\midrule\arrayrulecolor{black}
		$\U1_{R}$   & $0$ & $0$ & $0$ & $0$ & $-1$  & $+1$   & $-1$    & $+1$  \\
		\bottomrule
	\end{tabular}
	\caption{Particle content for \ac{DM} production.}
	\label{table:ModelDM}
\end{table}

\begin{figure}[t]
    \centering
\begin{subfigure}[t]{0.35\textwidth}
    \begin{tikzpicture}
    \begin{feynman}
      \vertex[dot,label =below:\(\,\alpha_m\)] (m) at (-1, 0) {};
      \vertex[dot,label =below:\(\beta_2\)] (n) at (1, 0) {};

      \vertex (a) at (-2,-1) {$\psi_{E_{i}}$};
      \vertex (b) at ( 2,-1) {$\psi_{\zeta_{\ell}}$};
      \vertex (c) at (-2, 1) {$\psi_{E^{C}_{j}}$};
      \vertex (d) at ( 2, 1) {$\psi_{\phi_{n}}$};
\coordinate[label=above:\(\phi_k\)] (n2)  at (0,0) {};

      \diagram* {
        (a)
        -- [fermion] (m)
        -- [fermion] (c),
        (b)
        -- [fermion] (n)
        -- [fermion] (d),
        (m)
        -- [scalar](n2)-- [scalar] (n),
      };
    \end{feynman}
  \end{tikzpicture}
        \caption{\ac{DM} production diagram.}
         \label{fig:DarkMatterProductionDiagram}
    \end{subfigure}\hspace{3em}
    \begin{subfigure}[t]{0.35\textwidth}
    \begin{tikzpicture}
    \begin{feynman}
      \vertex[dot,label =below:\(\sim \frac{\beta_2 \alpha_m}{m_\phi^2}\)] (m) at (0, 0) {};

      \vertex (a) at (-2,-1) {$\psi_{E_{i}}$};
      \vertex (b) at ( 2,-1) {$\psi_{\zeta_{\ell}}$};
      \vertex (c) at (-2, 1) {$\psi_{E^{C}_{j}}$};
      \vertex (d) at ( 2, 1) {$\psi_{\phi_{n}}$};

      \diagram* {
        (a)
        -- [fermion] (m)
        -- [fermion] (b),
        (c)
        -- [fermion] (m)
        -- [fermion] (d),
      };
    \end{feynman}
  \end{tikzpicture}
        \caption{Effective \ac{DM} production diagram}
        \label{fig:EffectiveDarkMatterProductionDiagram}
    \end{subfigure}
    \caption{(a). The indices $i,j,m$ indicate the three possible charged-lepton flavor states, while the indices $n,k,\ell$ indicate the three possible components of the triplet plus the non-trivial singlet of flavons and driving fields. \ac{DM} is identified as the lowest mass-state of $\psi_{\zeta_{\ell}},\psi_{\phi_{n}}$. The mediator is the scalar flavon $\phi_k$. (b) Since the effective mass of the flavon scalar is expected to be very big, $m_\phi \gg m_{E_{i}},m_{\psi_{\phi\zeta}}$, then the diagram shrinks to an effective $4-$fermion interaction. This leads to a preferred freeze-in scenario.}
\end{figure}
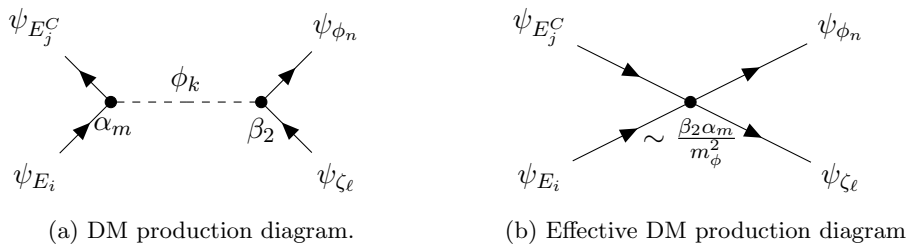

Interactions in \Cref{eq:L_CL,eq:L_phi} allow for processes as shown in the diagrams in \Cref{fig:DarkMatterProductionDiagram}. 
Furthermore, we also consider the scalar potential
\begin{equation}
    V = \frac{\partial W}
    	{\partial \varphi_{i} }K^{i\bar{\jmath}} \frac{\partial W^{*}}
    	{\partial \varphi_{\bar{\jmath}}^{*} } + \sum_{i=1,2,3} | \widetilde{m}_{\phi} \phi_{3,i} |^2 + | \widetilde{m}_{\phi} \phi_{1'}|^2\;,
    \label{eq:ScalarPotentialOfModelScalars}
\end{equation}
where we have added the soft-masses  $\widetilde{m}_{\phi}$ for the flavon scalars. We assume that  $\widetilde{m}_{\phi}$ is the same for the four scalars and should be of order of the \ac{SUSY} breaking scale. It has been shown in~\cite{Tanimoto:2021ehw} that for modular $A_4$ supersymmetric models, the \ac{SUSY} breaking scale is constrained to be above $6$\,TeV.
Therefore, we choose values of order $\widetilde{m}_{\phi} =20-1000$\,TeV. Furthermore, the $A$-term in \Cref{eq:SoftTerms} does not play a significant role in \ac{DM} production at tree-level since the production of \ac{DM} through flavon scalar annihilations are suppressed due to the low reheating temperature we consider; therefore, we also ignore the $A$-term in our analysis.

It turns out that the correct \ac{DM} relic abundance can be obtained with a freeze-in scenario~\cite{Hall:2009bx} in this model (cf.~\cite{Nomura:2024abu} where a freeze-out production mechanism is utilized). This can be seen as follows. First, note that the effective scalar flavon mass $m_\phi$ is obtained by diagonalizing the mass matrix obtained from \Cref{eq:ScalarPotentialOfModelScalars}. If we assume that $\widetilde{m}_{\phi} \gg   m_{E_{i}},m_{\psi_{\phi\zeta}}$, where $m_{E_{i}}, m_{\psi_{\phi\zeta}}=m_{\text{DM}}\sim \Lambda_\phi \beta_1$ represent the mass of the charged leptons and the \ac{DM} respectively, then the mediator mass ($m_{\phi} \sim \widetilde{m}_{\phi} + m_{\psi_{\phi\zeta}} $) is much larger than the \ac{DM} and the charged-lepton masses.  Therefore, the diagram  \Cref{fig:DarkMatterProductionDiagram} reduces to the effective $4-$fermion operator  as indicated in \Cref{fig:EffectiveDarkMatterProductionDiagram} with a coupling of $\sim \frac{\beta_2 \alpha_m}{m_\phi^2}$. In a freeze-out mechanism, if the rate $\left\langle \sigma v \right \rangle $ at which \ac{DM} is annihilated decreases, then the amount of  \ac{DM} relic abundance increases. Since we must require $\beta_2 \ll 4 \pi$ to retain perturbativity, and we expect $\alpha_1 = -6.97$, as explained at the end \Cref{Sec:FlavorFit}, then
\begin{equation}
\label{eq:averageCrossSec}
\left\langle \sigma v \right \rangle\sim \left|\frac{\beta_2 \alpha_m}{m_\phi^2} \right|^2 \ll \left|\frac{88}{m_\phi^2} \right|^2\;.
\end{equation}

Using \texttt{micrOMEGAs} 5.3.41~\cite{Belanger:2018ccd,Alguero:2022inz,Belanger:2021smw}, we find that for the chosen values of $\Lambda_{\phi}$ and $\widetilde{m}_{\phi}$, too much \ac{DM} is produced. Since increasing $m_{\phi}$ or lowering $\beta_2$ would only decrease $\left\langle \sigma v \right\rangle$, the \ac{DM} abundance can not be decreased to the observed \ac{DM} abundance in a freeze-out scenario.\footnote{If we had chosen high enough reheating temperatures, then the $A-$terms of \Cref{eq:SoftTerms} would have dominated the production of \ac{DM}. In this case, the connection between \ac{DM} and flavor parameters is weakened.} On the other hand, for a freeze-in scenario, we have the opposite behavior. Specifically, if $\left\langle \sigma v \right \rangle $ decreases, the amount of produced \ac{DM} also decreases. So, we can choose smaller $\beta_{2}$ or larger $m_{\phi}$ values to obtain the observed relic abundance of \ac{DM} in the Universe.

We now proceed to present the predictions of our model for the \ac{DM} abundance after performing a parameter scan. We use \texttt{micrOMEGAs} 5.3.41~\cite{Belanger:2018ccd,Alguero:2022inz,Belanger:2021smw} for the \ac{DM} abundance computation and FeynRules 2.0~\cite{Alloul:2013bka} to create the \texttt{CalcHEP}~\cite{Belyaev:2012qa} model files. As mentioned earlier, we assume $\tan \beta = 60 $ and a low reheating temperature of $T_{\text{R}} =150$\,GeV.

From our discussion we see that we have 4 free parameters to determine the \ac{DM} in our model:
\begin{enumerate}
\item the scalar flavon soft-mass $\widetilde{m}_{\phi}$,
\item the flavor breaking scale $\Lambda_{\phi}$,
\item the coupling $\beta_1$ in \Cref{eq:Wphi_model}, and
\item the coupling $\beta_4$ in \Cref{eq:Wphi_model}.
\end{enumerate}
We fix $10^{-6} \leq \beta_1, \beta_4 \leq 10^{-4}$ and  $20$\,TeV $ \leq \widetilde{m}_{\phi}, \, \Lambda_{\phi} \leq 1000$\,TeV.
These bounds for the couplings $\beta_1,\, \beta_4$ respect the perturbativity of all the couplings $\beta_i$ (cf.~\Cref{eq:couplings_FtermNumerical}).

\begin{figure}[t]
\centering
\begin{subfigure}[b]{0.45\textwidth}
         \centering
         \includegraphics[width=\textwidth]{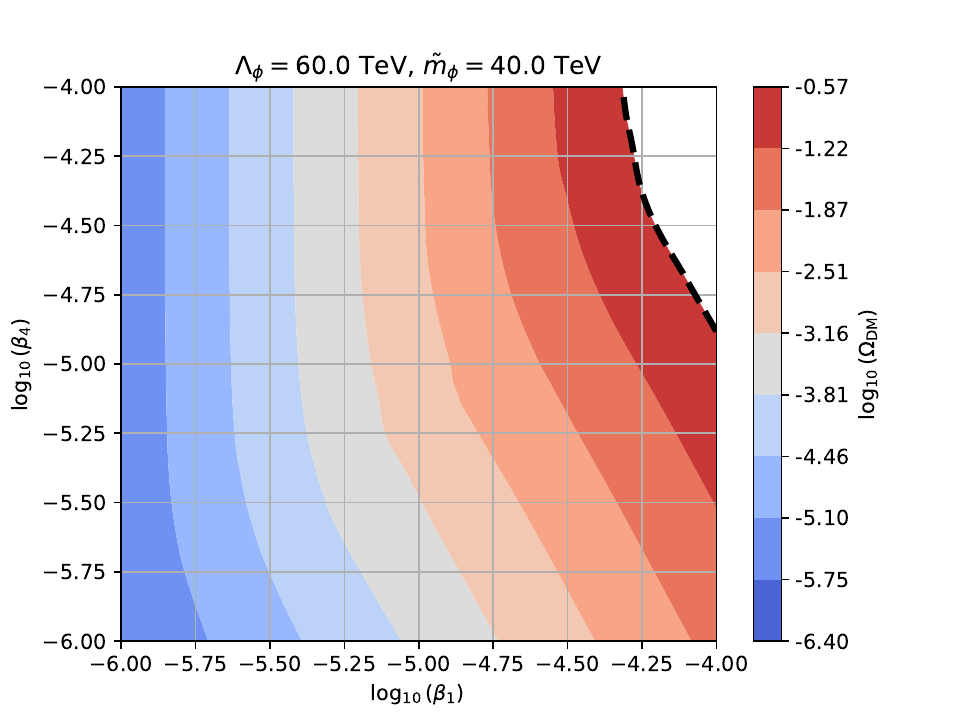}
         \caption{}
         \label{fig:LambdaphiBiggersm32}
     \end{subfigure}
\begin{subfigure}[b]{0.45\textwidth}
         \centering
         \includegraphics[width=\textwidth]{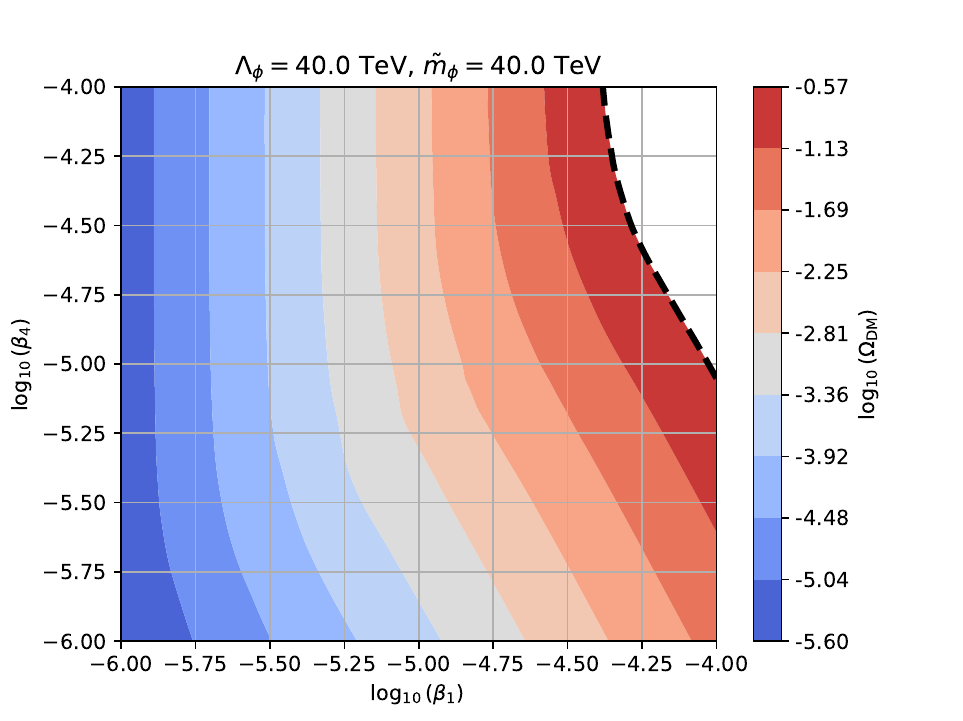}
         \caption{}
         \label{fig:LambdaphiEqualsm32}
     \end{subfigure}
\begin{subfigure}[b]{0.45\textwidth}
         \centering
         \includegraphics[width=\textwidth]{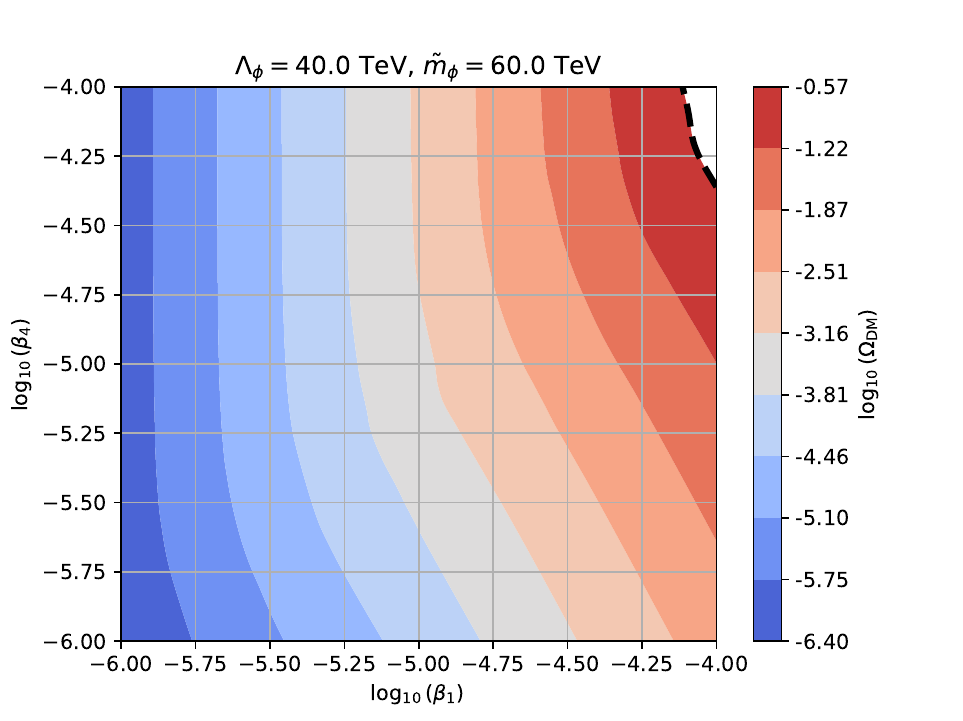}
         \caption{}
         \label{fig:LambdaphiLesssm32}
     \end{subfigure}
\caption{Predicted \ac{DM} relic abundance as a function of $\log_{10}\left( \beta_1 \right)$ and $\log_{10}\left( \beta_4 \right) $ with fixed values of $\widetilde{m}_{\phi}$ and $\Lambda_{\phi}$ at some benchmark values. We choose $T_{\text{R}} =150$\,GeV. The dashed black line denotes the relic \ac{DM} abundance $\Omega_{\text{DM}} = 0.265 $.}
\label{fig:Beta1VsBeta4}
\end{figure}

In \Cref{fig:Beta1VsBeta4} we show the prediction for \ac{DM} relic abundance as a function of $\beta_1$ and $\beta_4$, for $\Lambda_{\phi} > \widetilde{m}_{\phi}$ (\Cref{fig:LambdaphiBiggersm32}), $\widetilde{m}_{\phi} = \Lambda_{\phi}$ (\Cref{fig:LambdaphiEqualsm32}), and $\widetilde{m}_{\phi} < \Lambda_{\phi}$ (\Cref{fig:LambdaphiLesssm32}). The unshaded region indicates the excluded parameter space where too much \ac{DM} is produced. We see that all plots in \Cref{fig:Beta1VsBeta4} exhibit similar behavior. It is possible to have a coupling up to $\beta_4, \beta_1 = 10^{-4}$, but not at the same time. This is consistent with the fact that freeze-in normally requires small couplings~\cite{Hall:2009bx}. Furthermore, the abundance increases as we increase either $\beta_1$ or $\beta_4$, which is consistent with the fact that  the \ac{DM} relic abundance increases with $\left\langle \sigma v \right \rangle $.

\begin{figure}[t]
\centering
\begin{subfigure}[b]{0.45\textwidth}
         \centering
         \includegraphics[width=\textwidth]{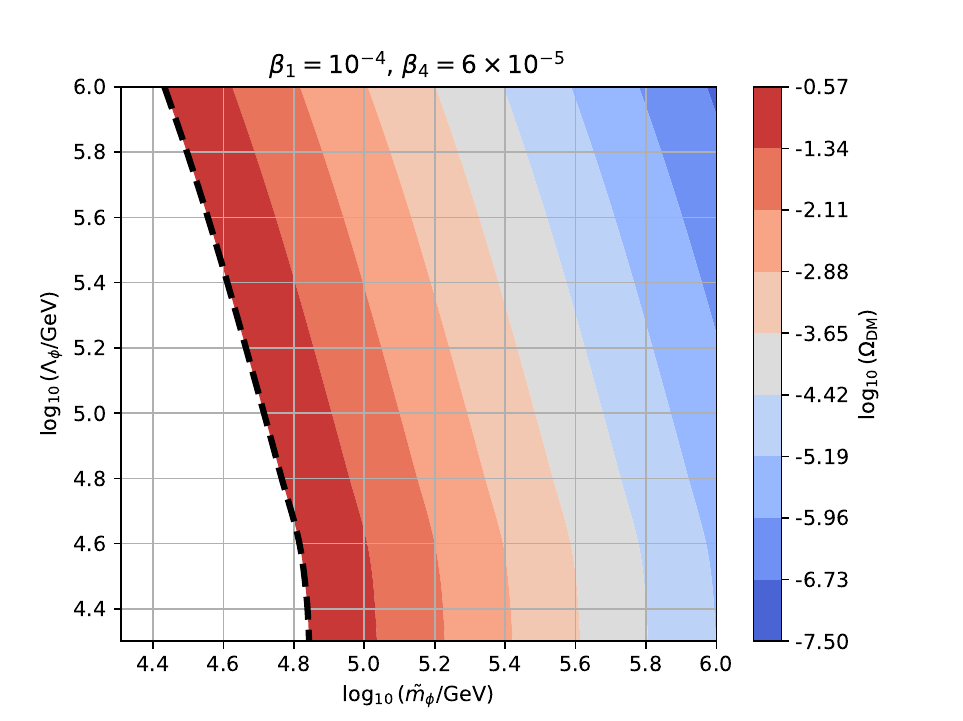}
         \caption{}
         \label{fig:Beta1BiggerBeta4}
     \end{subfigure}
\begin{subfigure}[b]{0.45\textwidth}
         \centering
         \includegraphics[width=\textwidth]{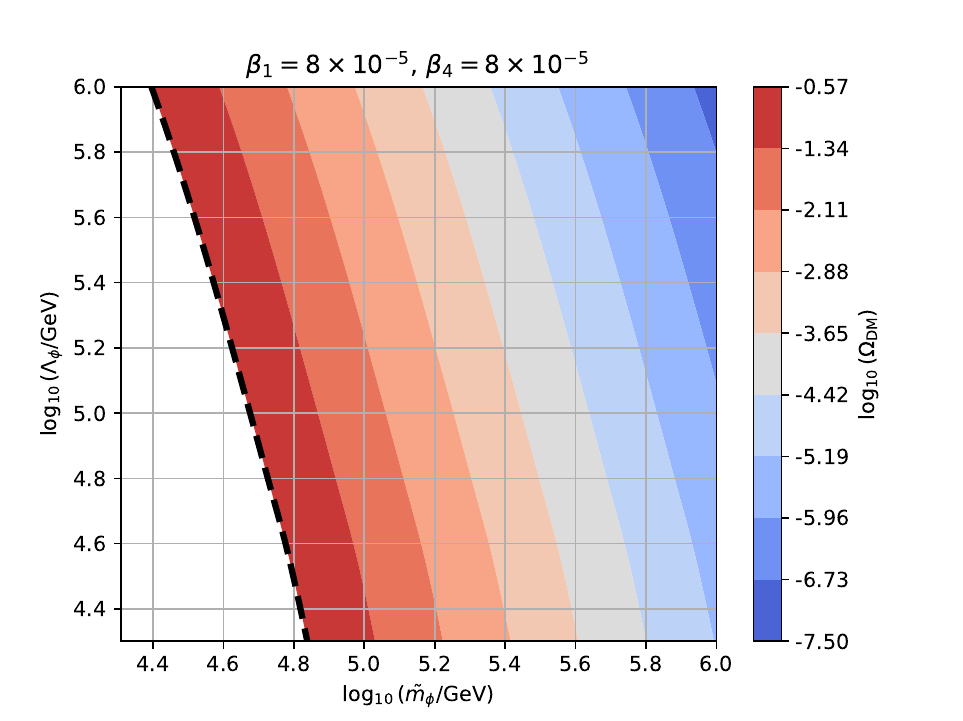}
         \caption{}
         \label{fig:Beta1EqualsBeta4}
     \end{subfigure}
\begin{subfigure}[b]{0.45\textwidth}
         \centering
         \includegraphics[width=\textwidth]{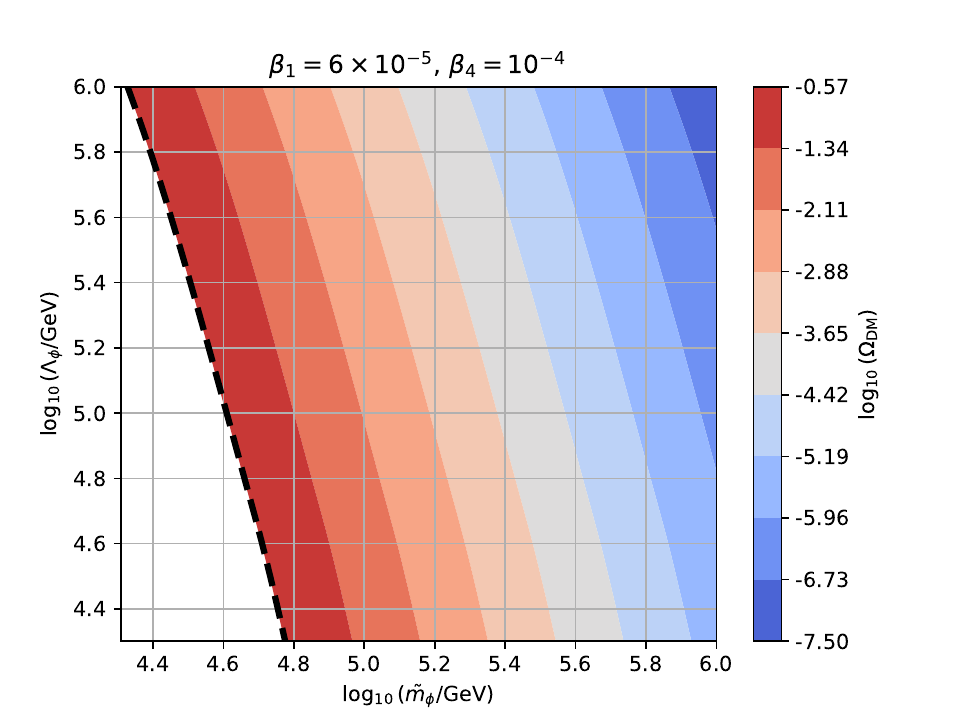}
         \caption{}
         \label{fig:Beta1LessBeta4}
     \end{subfigure}
\caption{Predicted \ac{DM} relic abundance as a function of $\log_{10}\left( \widetilde{m}_\phi / \text{GeV} \right)$ and $\log_{10}\left( \Lambda_\phi / \text{GeV} \right)$ with fixed values of $\beta_1$ and $\beta_4$ at some benchmark values. We choose $T_{\text{R}} =150$\,GeV. The dashed black line denotes the relic \ac{DM} abundance $\Omega_{\text{DM}} = 0.265 $.}
\label{fig:LambdaphiVsM32}
\end{figure}

\Cref{fig:LambdaphiVsM32} shows the predicted \ac{DM} relic abundance as a function of $\widetilde{m}_\phi$ and $\Lambda_\phi$ for $\beta_1 > \beta_4$ (\Cref{fig:Beta1BiggerBeta4}), $\beta_1 = \beta_4$ (\Cref{fig:Beta1EqualsBeta4}), and $\beta_1 < \beta_4 $ (\Cref{fig:Beta1LessBeta4}). The unshaded space represents the excluded parameter space. We observe a similar behavior for the three plots in \Cref{fig:LambdaphiVsM32}. The \ac{DM} abundance decreases if either $\widetilde{m}_\phi$ or $\Lambda_\phi$ grows. Furthermore, a soft-mass of $m_{\phi}= 20$\,TeV and a simultaneous flavon breaking-scale $\Lambda_{\phi}= 20$\,TeV is excluded in all cases for the chosen values of other parameters.

\begin{figure}[t]
\centering
\begin{subfigure}[b]{0.45\textwidth}
         \centering
         \includegraphics[width = \textwidth]{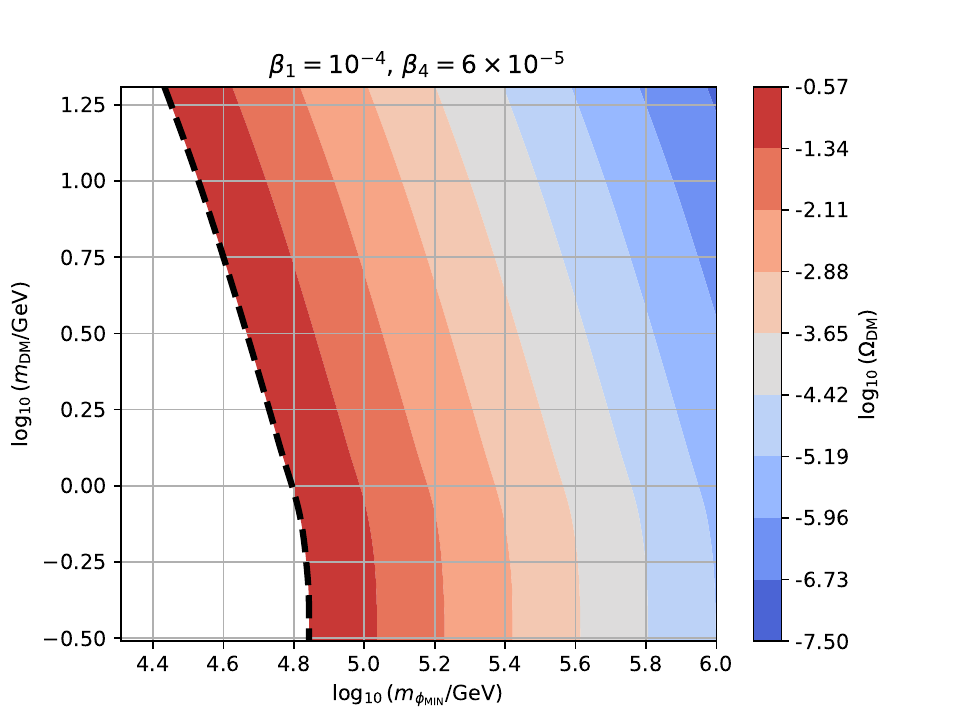}
         \caption{}
         \label{fig:Beta1BiggerBeta4Masses}
     \end{subfigure}
      \hfill
\begin{subfigure}[b]{0.45\textwidth}
         \centering
         \includegraphics[width = \textwidth]{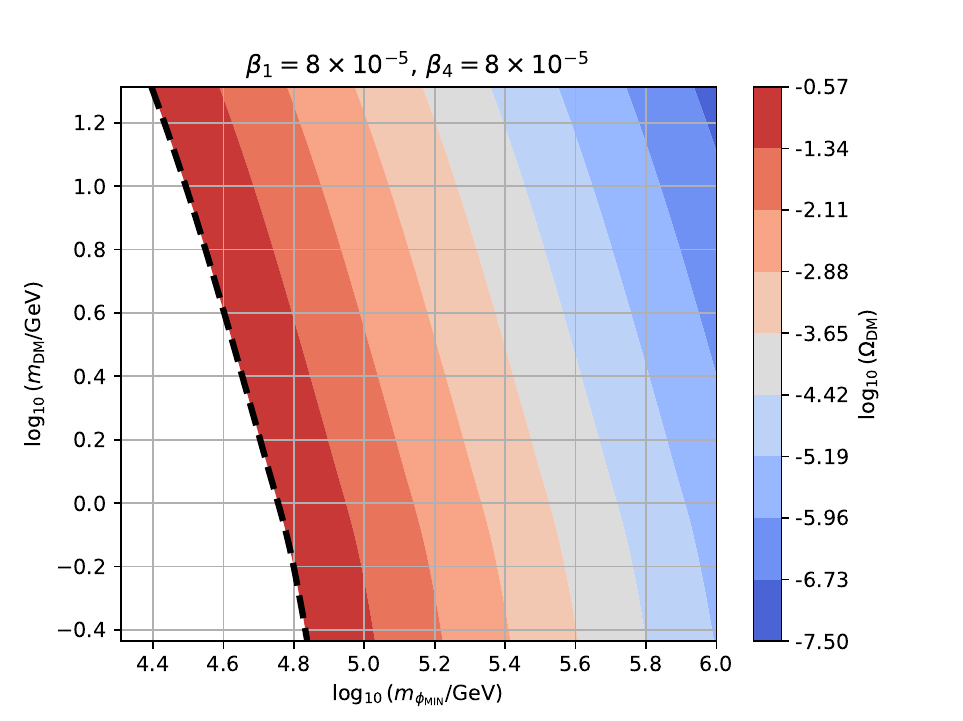}
         \caption{}
         \label{fig:Beta1EqualsBeta4Masses}
     \end{subfigure}
      \hfill
\begin{subfigure}[b]{0.45\textwidth}
         \centering
         \includegraphics[width = \textwidth]{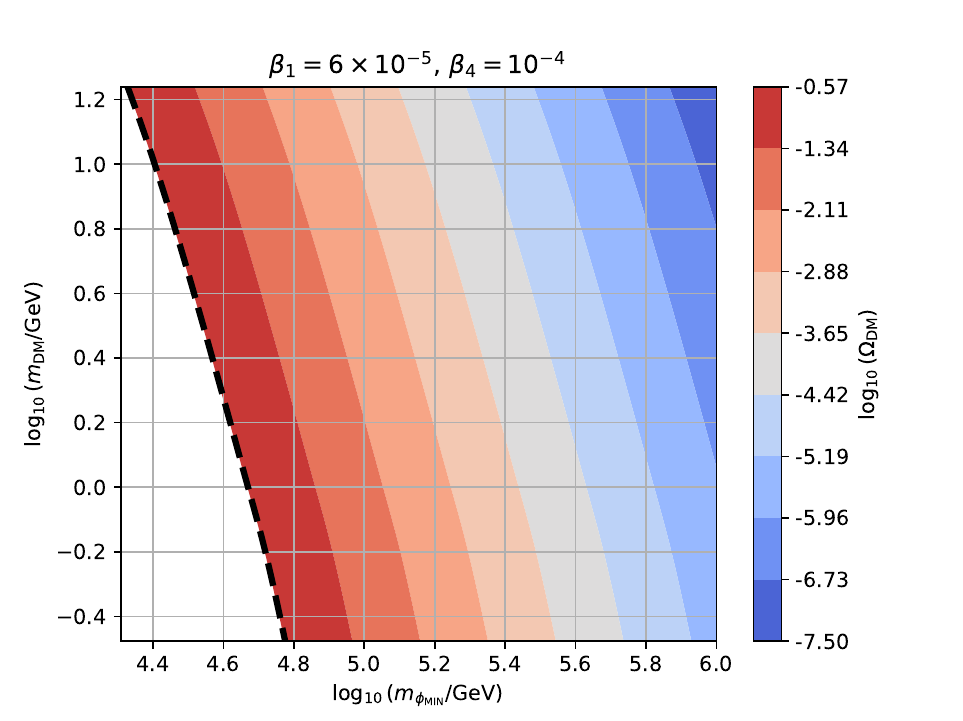}
         \caption{}
         \label{fig:Beta1LessBeta4Masses}
     \end{subfigure}
\caption{Predicted \ac{DM} relic abundance as a function of the minimum mediator scalar flavon mass $\log_{10}\left( m_{\phi_{\text{MIN}}} / \text{GeV} \right)$  and \ac{DM} mass $\log_{10}\left(  m_{\text{DM}} / \text{GeV} \right)$ for fixed $\beta_1$ and $\beta_4$  at some benchmark values. We fix $T_{\text{R}} =150\text{ GeV}$. The dashed black line denotes the relic \ac{DM} abundance $\Omega_{\text{DM}} = 0.265 $.}
\label{fig:MphiminVsMDM}
\end{figure}

Finally, in \Cref{fig:MphiminVsMDM} we show the correlation between the minimum scalar flavon mass $m_{\phi_{\text{MIN}}}$ and \ac{DM} mass $m_{\text{DM}}$ for fixed values of $\beta_1 > \beta_4$ (\Cref{fig:Beta1BiggerBeta4Masses}), $\beta_1 = \beta_4 $ (\Cref{fig:Beta1EqualsBeta4Masses}), and $\beta_1 < \beta_4 $ (\Cref{fig:Beta1LessBeta4Masses}). By comparing \Cref{fig:MphiminVsMDM,fig:LambdaphiVsM32}, we find that for a given value of the soft-mass $\widetilde{m}_{\phi}$, the minimum scalar flavon mass $m_{\phi_{\text{MIN}}}$ constrained by the \ac{DM} relic abundance can be derived. We see that the scalar flavon mass can be as low as $\sim63$\,TeV for all cases illustrated in \Cref{fig:MphiminVsMDM}. Moreover, the \ac{DM} mass can be as low as $\sim16$\,GeV for $\beta_1 = 6\times 10^{-5}$ and $\beta_4 =10^{-4}$. Note that the big mass splitting between the mediator and the \ac{DM}, as seen in \Cref{fig:MphiminVsMDM}, is a consequence from the big soft-mass contribution from supersymmetry breaking to the mediator mass (see discussion before \Cref{eq:averageCrossSec}). A mechanism for supersymmetry breakdown is beyond the scope of this work and must be explored elsewhere.

To conclude this section, let us comment on the bounds from direct detection for this model. The energy-independent cross section for the scattering process $\psi\,e^{-}\to{}\psi\,e^{-}$ between a \ac{DM} particle $\psi$ and the electron $e^{-}$ is given by (cf.~\cite[Equations (9) and (10)]{Bardhan:2022bdg})
\begin{equation}
\label{eq:SigmaApprox}
   \bar{\sigma}_{e\psi} ~=~ \frac{\mu_{e\,\psi}^2}{\pi}\frac{\alpha^2 \beta^2}{(q_{\text{ref}}^2-m_\phi^2)^2}\, ,
\end{equation}
where $q_{\text{ref}} = \alpha_{EM} m_e$ is the reference momentum (with the electromagnetic fine-structure constant $\alpha_{EM} \approx 1 / 137$ and the electron mass $m_e$), $\mu_{e\,\psi}$ denotes the reduced mass of the \ac{DM} candidate and the electron, $\alpha$ and $\beta$ are respectively the mediator--electron and mediator--\ac{DM} couplings, and $m_\phi$ is the mediator mass. Assuming $m_\psi = 1$\,GeV, $\alpha = 10^{-3}$, $\beta = 10^{-4}$ and $m_\phi = 20$\,TeV (consistent with couplings shown in \Cref{eq:BestFitFlavor} and \Cref{fig:Beta1VsBeta4,fig:LambdaphiVsM32,fig:MphiminVsMDM}), we obtain
\begin{equation}
\label{eq:ModelResultingSigma}
   \bar{\sigma}_{e\psi} \sim 10^{-66}\, \text{cm}^{2}\,,
\end{equation}
which is unfortunately found below the direct detection bounds from XENON1T~\cite{XENON:2019gfn}, for a \ac{DM} mass of $1$\,GeV. It is known that the detection bounds can change depending on the energy-dependence of the cross section, or on the form of the so-called \ac{DM} form factor $F_{\text{DM}}(q)$. However, we do not expect the energy-independent cross section in \Cref{eq:ModelResultingSigma} to significantly increase to become experimentally detectable (see~\cite[Figure 3]{DAMIC:2019dcn}). This implies that the DM features of our model require new experimental settings.

\section{Conclusion}
\label{sec:SummaryAndOutlook}

We constructed a flavor model based on the finite modular group $\Gamma_3\cong A_{4}$ that simultaneously explains the flavor parameters in the lepton sector and accounts for the observed \ac{DM} relic abundance in the Universe. The 12 lepton flavor parameters are determined by 8 real parameters: the two components of the modulus \ac{VEV}, the four flavon \ac{VEV}s, $\left\langle \phi_{i} \right \rangle$, and two dimensionful parameters that set the mass scales of charged-lepton and neutrino masses. We identify a \ac{DM} candidate composed of the fermionic parts of the flavon superfields and the driving superfields. The mediator is the scalar part of the flavon superfield which interacts with the charged-lepton sector of the \ac{SM}. We obtain a good fit to the lepton flavor parameters (3 charged lepton masses, 3 mixing angles, 1 \CP phase, 2 neutrino squared mass differences) with $\chi^2 = 0.08$ for an inverted-hierarchy neutrino spectrum. The lepton flavor fit fixes the couplings of the charged leptons to the \ac{DM} mediator as well as the flavon \ac{VEV}s $\left\langle \phi_{i} \right \rangle$.
These \ac{VEV}s satisfy the $F$-term equations to retain supersymmetry at high energies, determining thereby the coupling between our \ac{DM} candidate and the mediator. Interestingly, our model exhibits 4 additional degrees of freedom that are left free and serve to achieve a \ac{DM} relic abundance which  does not exceed the observed value $\Omega_{\text{CDM}} = 0.265$. These parameters are 2 dimensionless couplings $\beta_1, \beta_4$, the flavor breaking scale $\Lambda_\phi$, and the soft-mass for the flavon $\widetilde{m}_{\phi}$. We find that if the mediator mass is assumed to be much larger than the \ac{DM} and the charged-lepton masses, then the appropriate \ac{DM} production mechanism is freeze-in rather than freeze-out. We observe that a viable \ac{DM} relic abundance can be generated in regions of the parameter space constrained by $10^{-6} \leq \beta_1,\beta_4 \leq 10^{-4}$,  $20$\,TeV $ \leq \widetilde{m}_{\phi}, \, \Lambda_{\phi} \leq 1000$\,TeV, $\tan\beta = 60$, and $T_R = 150$\,GeV.

Although some amount of tuning is necessary in our model to identify the best parameter values, we point out that it is the choice of charges and modular weights that render the right flavon superpotential, which in turn delivers the alignment of the flavon \ac{VEV}s $\left\langle \phi_{i} \right \rangle$. Further, the phenomenologically viable value of the modulus \ac{VEV} $\left\langle \tau \right \rangle $ might be achieved through mechanisms as in~\cite{Knapp-Perez:2023nty}. The flavor structure of our model, that includes the flavon superpotential, determines the properties of our \ac{DM} candidate. In particular, a different choice of flavor parameters in~\Cref{eq:BestFitFlavor,eq:couplings_FtermNumerical}, which set the flavor predictions, strongly influences the mass and production (e.g.\ freeze-in vs.\ freeze-out) of \ac{DM} in our model.

As a feasible outlook of our findings, it would  be interesting to study the possibility of applying this new scenario in top-down models such as~\cite{Baur:2022hma}. Furthermore, as the flavor sector of our model only accounts for leptons, it should be extended to the quark sector as a first direct step. In addition, in our scenario we expect the \ac{DM}-nucleon cross section to be too small to be directly compared with LUX~\cite{LUX:2018akb}, DEAP-3600~\cite{DEAP:2019yzn}, PandaX-II~\cite{PandaX-II:2017hlx}, DarkSide~\cite{DarkSide:2018bpj}, EDELWEISS~\cite{EDELWEISS:2019vjv} and any other currently envisaged experiments of this type. One should hence explore additional indirect evidence of our proposal. We leave these intriguing questions for upcoming work.

\subsection*{Acknowledgments}
The work of M.-C.C. and V.K.-P. was partially supported by the U.S. National Science Foundation under Grant No. PHY-1915005. 
The work of S.R.-S. was partly supported by UNAM-PAPIIT grant IN113223 and Marcos Moshinsky Foundation. This work was also supported by UC-MEXUS-CONACyT grant No.\ CN-20-38.
V.K.-P.  would like to thank Kevork N. Abazajian, Max Fieg, Xueqi Li, Xiang-Gan Liu, Gopolang Mohlabeng, Michael Ratz and Mi\v{s}a Toman for fruitful discussions. M-C.C.and V.K.-P. would also like to thank Instituto de F\'isica, UNAM, for the hospitality during their visit. A.B. would like to thank the Department of Physics and Astronomy at UCI for the hospitality during his visit. V.K.-P. also thanks the opportunity to use the cluster at Instituto de F\'isica, UNAM. This work utilized the infrastructure for high-performance and high-throughput computing, research data storage and analysis, and scientific software tool integration built, operated, and updated by the Research Cyberinfrastructure Center (RCIC) at the University of California, Irvine (UCI). The RCIC provides cluster-based systems, application software, and scalable storage to directly support the UCI research community. https://rcic.uci.edu

\newpage
\providecommand{\bysame}{\leavevmode\hbox to3em{\hrulefill}\thinspace}
\frenchspacing
\newcommand{\origttfamily}{}
\let\origttfamily=\ttfamily
\renewcommand{\ttfamily}{\origttfamily \hyphenchar\font=`\-}

\begin{acronym}
\acro{DM}{dark matter}
\acro{SUSY}{supersymmetry}
\acro{VEV}{vacuum expectation value}
\acro{EFT}{effective field theory}
\acro{MFV}{minimal flavor violation}
\acro{SM}{Standard Model}
\acro{EW}{Electroweak}
\end{acronym}

\end{document}